\documentclass[a4paper]{cas-dc}
\usepackage[authoryear,longnamesfirst]{natbib}
\usepackage{hyperref}
\usepackage{graphicx}
\usepackage{multirow}
\usepackage{tabularx}
\usepackage{array}
\usepackage{longtable}
\usepackage{setspace}
\usepackage{xcolor}

\begin{document}

\let\WriteBookmarks\relax
\def\floatpagepagefraction{1}
\def\textpagefraction{.001}

\shorttitle{A Narrative Review of Image Processing Techniques Related to Prostate Ultrasound}

\shortauthors{H. Wang et al} 

\title [mode = title]{A Narrative Review of Image Processing Techniques Related to Prostate Ultrasound} 

\tnotetext[1]{The version of record of this contribution is accepted and published in \textit{Ultrasound in Medicine \& Biology}.}

%

\author[1,2]{Haiqiao Wang}
\author[1]{Hong Wu}
\author[1]{Zhuoyuan Wang}
\author[1]{Peiyan Yue}
\author[1]{Dong Ni}
\author[2]{Pheng-Ann Heng}
\author[1]{Yi Wang\corref{cor1}}[orcid=0000-0002-8428-288X]
\cortext[cor1]{Corresponding author: Yi Wang (onewang@szu.edu.cn)}

\address[1]{Medical UltraSound Image Computing (MUSIC) Lab, Smart Medical Imaging, Learning and Engineering (SMILE) Lab, School of Biomedical Engineering, Shenzhen University Medical School, Shenzhen University, Shenzhen, China}
\address[2]{Department of Computer Science and Engineering, The Chinese University of Hong Kong, Hong Kong, China}


\begin{abstract}
Prostate cancer (PCa) poses a significant threat to men's health, with early diagnosis being crucial for improving prognosis and reducing mortality rates.
Transrectal ultrasound (TRUS) plays a vital role in the diagnosis and image-guided intervention of PCa.
To facilitate physicians with more accurate and efficient computer-assisted diagnosis and interventions, many image processing algorithms in TRUS have been proposed and achieved state-of-the-art performance in several tasks, including prostate gland segmentation, prostate image registration, PCa classification and detection, and interventional needle detection.
The rapid development of these algorithms over the past two decades necessitates a comprehensive summary.
In consequence, this survey provides a narrative review of this field, outlining the evolution of image processing methods in the context of TRUS image analysis and meanwhile highlighting their relevant contributions.
Furthermore, this survey discusses current challenges, and suggests future research directions to possibly advance this field further.
\end{abstract}


\begin{keywords}
Transrectal ultrasound \sep
Prostate cancer \sep
Medical image processing\sep
Deep learning \sep
Machine learning \sep
Medical image segmentation \sep
Medical image registration\sep
Classification \sep
Computer assisted detection\sep
Computer assisted diagnosis
\end{keywords}

\maketitle

\section{Introduction}
Prostate cancer (PCa) is a malignant disease of the prostate gland.
According to~\citep{siegel2023cancer, xia2022cancer}, in 2022, the number of PCa patients ranked second among male cancer patients worldwide,
accounting for 14.2\%, while the number of deaths from PCa ranked fifth among male cancer patients, accounting for 7.3\%.
Early diagnosis and treatment is the crucial key to the cure of progressive PCa.
If this can not be carried out effectively, PCa can invade nearby tissues and spread to other parts of the body, such as bones, seminal vesicles, and the rectum~\citep{rawla2019epidemiology}.
Such spread not only severely affects the patient's quality of life but also increases the risk of death.
The prevalence of subclinical and undiagnosed PCa is high among elderly men, and the incidence of asymptomatic PCa is also significant.
Approximately half of patients with low-risk PCa (Gleason score $\le$6 / Grade Group 1) ultimately require active treatment.
For low-risk PCa, more active monitoring is an effective approach~\citep{semsarian2023low}.
Therefore, early diagnosis for PCa is crucial for improving the success rate of treatment and reducing mortality.

The diagnostic procedure for PCa requires the comprehensive use of various detection tools to ensure accuracy.
Firstly, digital rectal examination (DRE) is a traditional method for preliminary assessment of the prostate condition, where the size, shape, and texture of the prostate are checked through rectal palpation~\citep{naji2018digital}.
Secondly, prostate specific antigen (PSA) blood testing is an important screening tool that assesses the risk of prostate disease by measuring the level of PSA in the blood, despite the possibility of its elevation due to non-cancerous reasons~\citep{lilja2008prostate}.
Additionally, imaging examinations, including ultrasound, computed tomography (CT) and magnetic resonance imaging (MRI), provide physicians with more detailed views of the prostate structure~\citep{hricak2007imaging}, with multi-parametric MRI (mpMRI) offering higher resolution images that help to more accurately locate tumors.
Ultimately, prostate biopsy is a key step in confirming prostate cancer, determining the nature of the tumor through pathological analysis.
However, early cancerous lesions are often small and difficult to accurately locate and obtain sufficient samples in one biopsy, usually requiring multiple attempts.
Multiple biopsies not only increase patient discomfort but can also lead to complications such as prostatitis or urinary tract infections~\citep{borghesi2017complications}.
To reduce this over-diagnosis phenomenon, additional imaging guidance for targeted prostate biopsy is essential~\citep{brown2015recent}, such as MRI and ultrasound.

Transrectal ultrasound (TRUS) guided prostate biopsy is the standard method for the diagnosis of PCa, which is widely used in clinical practice~\citep{ahmed2017diagnostic}.
TRUS scans the prostate through the rectal wall, generating images of the prostate, including B-mode, color Doppler ultrasound (CDUS), contrast-enhanced ultrasound (CEUS), 3D imaging, and elastography, to meet various clinical needs.
The real-time B-mode provides two-dimensional images commonly used for the diagnosis of benign prostate hyperplasia (BPH) and prostatitis, and intraoperative guidance.
While 3D TRUS generates comprehensive 3D volumetric information that is particularly important for preoperative planning.
Furthermore, micro-ultrasound, operates at a higher frequency than conventional ultrasound, providing higher resolution imaging.
However, this technology is not yet widely available, and there is limited imaging processing research related to it.
CEUS and CDUS are used to identify suspicious regions based on the assumption that microvascular density increases in PCa due to its association with angiogenesis.
These imaging techniques can easily detect areas with dense blood vessels, assisting clinicians in quickly identifying suspicious vascular regions~\citep{frauscher2002comparison}.
Elastography, on the other hand, relies on the assumption that PCa regions tend to exhibit increased stiffness, often due to higher cell density, which reduces tissue elasticity.
These stiffened areas can be detected by strain elastography (SE) or shear wave elastography (SWE), enabling clinicians to locate potential regions of concern based on increased tissue hardness.
However, using these two modalities alone to localize PCa can lead to false positives.
Prostatitis and BPH can also cause increased vascularization, while benign nodules and prostate fibrosis can result in tissue stiffening.
Therefore, it is essential for clinicians to integrate information from multiple modalities to accurately localize PCa~\citep{oladimeji2024predictive}.

Here are several key clinical tasks where TRUS is needed as imaging tool for the diagnosis and treatment of PCa:
1) \textit{\textbf{PCa classification and detection}}:
TRUS can help physicians identify different types of PCa, including localized and invasive PCa.
Through TRUS images, physicians can roughly assess the size, location, and invasion of tumors into surrounding tissues.
2) \textit{\textbf{Prostate gland segmentation}}:
accurate prostate boundary delineation from TRUS images can provide the target area for interventions, which facilitates physicians with effective treatment planning, biopsy needle placement, and brachytherapy.
3) \textit{\textbf{Image guidance for prostate intervention}}:
TRUS is the routine imaging tool for image-guided biopsy and therapy of PCa, due to its real-time advantage.
Moreover, other preoperative imaging modalities such as MRI can be registered/fused with TRUS to provide more comprehensive visual information,  thereby improving the efficacy of the intervention.
4) \textit{\textbf{Intraoperative needle localization}}:
during the procedure of biopsy, brachytherapy, cryotherapy and so on, 
TRUS is used to monitor the position of the biopsy or treatment needle, ensuring the accuracy and safety of the intervention.

Although TRUS has been widely used for the imaging of prostate, there are still limitations in practical applications.
The image quality may be degraded by speckle noise and shadow, leading to blurred details and artifacts.
Especially in the apex and base regions of the prostate, the gland's contour in TRUS is often unclear, increasing the complexity of diagnosis.
In addition, the interpretation of TRUS depends on the experience and skill of the physicians, leading to subjectivity.
Moreover, the modalities and dimensions of TRUS are various, making it difficult for physicians to quickly obtain potentially useful information during diagnosis and treatment.
Therefore, applying image processing technology to analyze TRUS is with clinical significance.

\begin{figure*}[t]
	\centering
	\includegraphics[width=\textwidth]{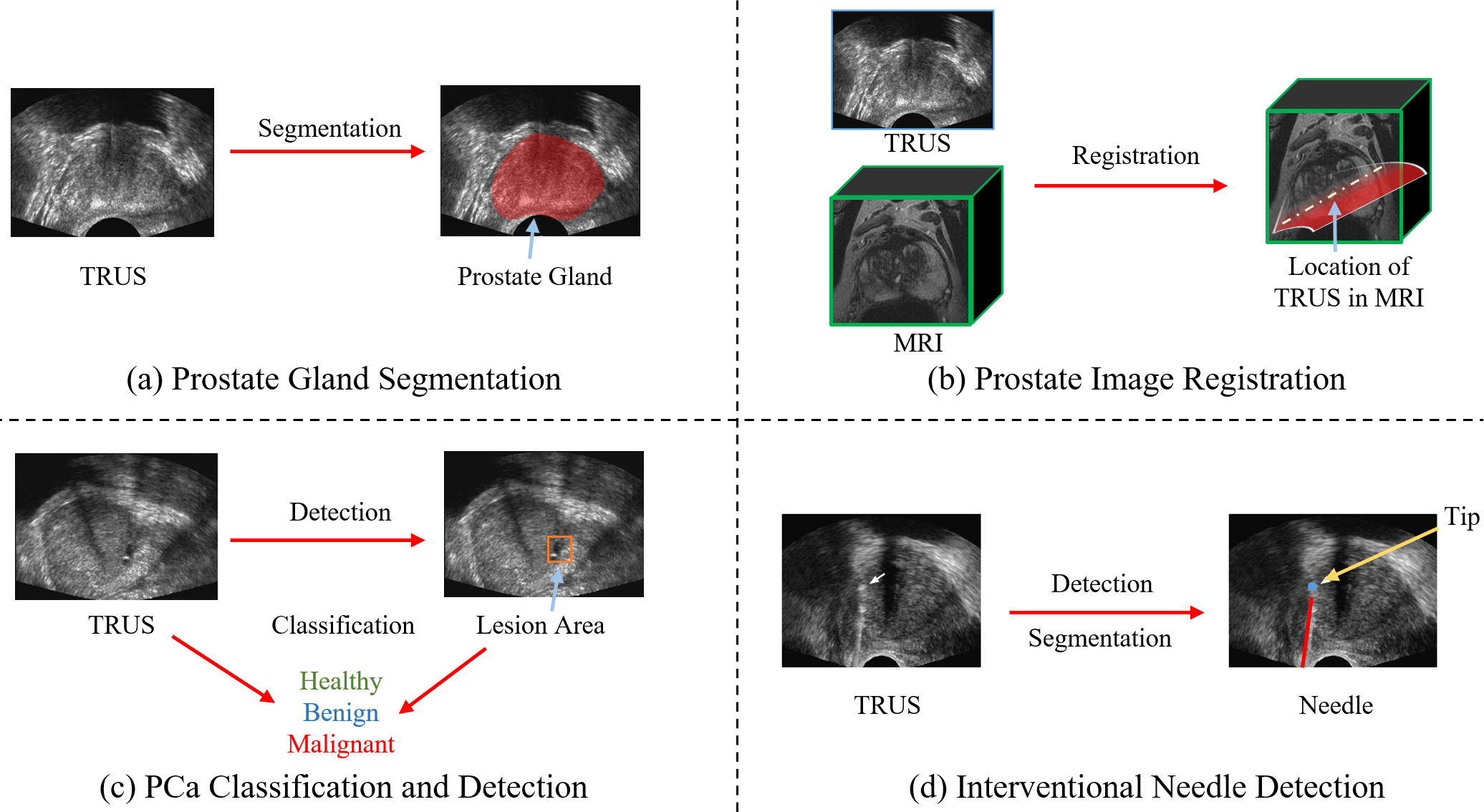}
	\caption{Illustration of conventional image processing tasks on transrectal ultrasound (TRUS), including (a) prostate gland segmentation, (b) prostate image registration, (c) prostate cancer (PCa) classification and detection, and (d) interventional needle detection.}
	\label{fig:task}
\end{figure*}

The development of image processing technology has greatly promoted the progress of the medical imaging field \citep{shen2017deep, litjens2017survey, 824822}, especially playing an important role in computer assisted diagnosis (CAD) and computer assisted intervention (CAI).
With the help of advanced image processing methods, it is beneficial for the physicians to make more accurate and efficient clinical decisions.
In the field of TRUS image processing, lots of relevant methods have been developed and applied in different tasks, mainly including prostate gland segmentation, prostate image registration, PCa classification and detection, and interventional needle detection,
as illustrated in Fig.~\ref{fig:task}.

This survey summarizes the related image processing methods in TRUS over the past 20 years, including gland segmentation, image registration, PCa classification and detection, and interventional needle detection.
We classify and summarize these methods according to modality, dimension, and technical route, presenting the similarities and differences of these methods.
We also list the primary evaluation metrics for different tasks, as well as the results of representative methods under their datasets and metrics, making a rough comparison of different methods.
Finally, we also summarize these methods to propose current challenges faced by each task and possible future research directions.

The organization of this survey is as follows.
In Section~\ref{sec:task}, we define each task of TRUS image processing, and list the definitions and formulas of their primary evaluation metrics.
Section~\ref{sec:method} is the main part of this survey,
where we review and summarize the methods of prostate gland segmentation, prostate image registration, PCa classification and detection, and needle detection in Section~\ref{subsec:seg}, \ref{subsec:reg}, \ref{subsec:cla} and \ref{subsec:det}, respectively.
In Section~\ref{sec:discussion}, we discuss and analyze existing challenges and suggest future research directions for these four tasks.
A summary of the entire survey is provided in Section~\ref{sec:conclusion}.

\section{Task Definition}
\label{sec:task}
\subsection{Prostate Gland Segmentation}
Image segmentation is one of the most common task in the field of medical image processing.
The purpose of prostate segmentation is to accurately obtain the contour/region of the prostate gland, see Fig.~\ref{fig:task}(a).
Prostate segmentation plays a crucial role in different stages of the clinical decision-making process.
For example, prostate volume, which can be directly determined through prostate segmentation, aids in the diagnosis of BPH.
Moreover, the boundary information of the prostate is important and useful in various treatment scenarios, such as prostate brachytherapy, high-intensity focused ultrasound (HIFU) therapy, cryotherapy, and transurethral microwave therapy.
In addition, prostate segmentation also facilitates other image processing tasks, such as surface-based multi-modality image registration.
To achieve prostate segmentation, lots of shape-/region-based methods, as well as conventional machine learning methods and advanced deep learning methods have been developed.
We classify and summarize these segmentation methods in Section~\ref{subsec:seg}, according to TRUS modality, dimension, and technical route.

\subsubsection{Evaluation Metrics}
The evaluation of segmentation accuracy can be divided into qualitative and quantitative evaluations.
For qualitative evaluation, the resulting contours are visually compared to the ground truth.
For quantitative evaluation, the numerical calculation involves measuring the similarity between the obtained contours and the ground truth.
The commonly used quantitative evaluation metrics consist of region-based metrics, boundary-based metrics, and classification-based metrics.
Since traditional segmentation methods often rely on contour constraints, they always employ boundary-based metrics for the evaluation.
In contrast, recent deep learning methods usually use region-based metrics.

Dice similarity coefficient (DSC) is currently the most popular region-based evaluation metric for the segmentation task,
which measures the overlap between the segmented result and the ground truth:
\begin{equation}
\label{formula:DSC}
\text{DSC} = \frac{2 \times TP}{FP + FN + 2 \times TP},
\end{equation}
where TP, FP, FN denotes true positive, false positive, and false negative regions, respectively.

Intersection over union (IoU) is another commonly used region-based metric, also known as the Jaccard coefficient.
It is defined as follows:
\begin{equation}
\label{formula:Jaccard}
\text{IoU}= \text{Jaccard} = \frac{TP}{TP + FP + FN}.
\end{equation}
Larger values of DSC and IoU indicate better segmentation accuracy.






As for the boundary-based metrics, mean surface distance (MSD), mean absolute distance (MAD), and average symmetric surface distance (ASD) are often employed to measure the similarity between the segmented boundary (in 2D)/surface (in 3D) and the ground truth:
\begin{equation}
\label{formula:MSD}
\text{MSD} = \frac{1}{N} \sum_{i=1}^{N} d_{i},
\end{equation}

\begin{equation}
\label{formula:MAD}
\text{MAD} = \frac{1}{N}\sum^N_{i=1}|d_i|,
\end{equation}

\begin{equation}
\label{formula:ASSD}
\text{ASD} = \frac{1}{N}\sum^{N}_{i=1}(|d_i|+|d_{i}'|),
\end{equation}
where $N$ is the total number of points from the segmented boundary/surface.
$d_i$ represents the shortest distance from the i-th point on the segmentation result to the ground truth,
while $d_{i}'$ represents the shortest distance from the i-th point on the ground truth to the segmentation result.
It can be seen that MSD, MAD, and ASD measures the average distance, mean absolute distance, and average symmetric distance, respectively.

In contrast to MSD, MAD, and ASD, maximum distance (MAXD) and Hausdorff distance (HD) are used to represent the dissimilarity between the segmented boundary/surface and the ground truth:
\begin{equation}
\label{formula:MAXD}
\text{MAXD} = \max|d_i|,
\end{equation}

\begin{equation}
\label{formula:HD}
\text{HD} = \max\left(\max|d_i|, \max|d_{i}'|\right).
\end{equation}
MAXD measures the maximum distance between each point from the segmentation result and its corresponding point from the ground truth.
HD measures the maximum distance between the segmentation result and ground truth, and with 95HD denoting the 95$th$ percentile of HD.
For all aforementioned boundary-based metrics, smaller values denotes better segmentation accuracy.

%


In addition to region-/boundary-based metrics, point-wise classification-based metrics can also be applied to measure segmentation accuracy.
Accuracy (ACC) represents the proportion of correctly predicted pixels to the total number of pixels in the segmentation result:
\begin{equation}
\label{formula:ACC}
\text{ACC} = \frac{TP + TN}{TP + TN + FP + FN},
\end{equation}
where TN denotes true negative.
Sensitivity and Recall indicate the proportion of all positive pixels that are correctly segmented as true positive:
\begin{equation} 
\label{formula:Sen}
\text{Sensitivity} = \text{Recall} = \frac{TP}{TP + FN}.
\end{equation}
Precision indicates the proportion of all predicted positive pixels that are TP pixels:
\begin{equation}
\label{formula:Precision}
\text{Precision} = \frac{TP}{TP + FP}.
\end{equation}
There is usually a trade-off between precision and sensitivity. 
Specificity is the proportion of all negative pixels that are correctly segmented as true negative:
\begin{equation}
\label{formula:Spe}
\text{Specificity} = \frac{TN}{TN + FP}.
\end{equation}
Larger values of these classification-based metrics indicate better segmentation.

%

\subsection{Prostate Image Registration}
Image registration is the process of aligning two or more images acquired at varying times, viewpoints, or by different sensors.
Taking registering a pair of images as an example, one is denoted as fixed image and the other as moving image.
The objective is to estimate the optimal deformation field between the fixed and moving images, thus the warped moving image can be matched with the fixed image, enabling the alignment of the regions of interest.
Prostate image registration plays a crucial role in assisting surgeons during preoperative planning and intraoperative surgery, with TRUS being a frequently utilized modality in this context.
Its purpose is to provide surgeons with more complementary and valuable information.
This task often involves analyzing multi-modality images (e.g., MR to TRUS, see Fig.~\ref{fig:task}(b)), thus remaining challenging in clinical practice.
We classify and summarize relevant registration methods in Section~\ref{subsec:reg}, according to modality, dimension, deformation type and technical route.

\subsubsection{Evaluation Metrics}
Considering the purpose of image registration, the numerical evaluation metrics are mainly aimed at quantifying the similarity of corresponding regions between the fixed image and the warped moving image.
Therefore the metrics like DSC, HD, MAD, and MAXD can also be used to evaluate the registration performance, by calculating the overlap of corresponding regions or the distance of corresponding boundaries/surfaces.

Target registration error (TRE) is one of the most commonly used metrics in medical image registration.
It involves measuring the Euclidean distance of the manually identified corresponding landmarks in the fixed and moving images:
\begin{equation}
\label{formula:TRE}
\text{TRE} = \frac{1}{N}\sum_{i=1}^{N}\left \|L_{I_f}^i , L_{I_m}^i\circ\phi \right\| _2,
\end{equation}
where $L$ denotes the landmark set, and $N$ is the total number of corresponding landmark pair in $L$.
$I_f$ and $I_m$ are the fixed and moving images, respectively.
$\phi$ is the estimated deformation field, and $\circ$ denotes the warping operation.
$\left \| \cdot \right\| _2$ is the L2 norm.

Surface registration error (SRE) has the similar computing method with TRE, but the difference is SRE measures the similarity of surface points between fixed and moving images:
\begin{equation}
\label{formula:SRE}
\text{SRE} = \frac{1}{N}\sum_{i=1}^{N}\left \|S_{I_f}^i , S_{I_m}^i\circ\phi \right\| _2,
\end{equation}
where $S$ denotes the prostate surface point set, and $N$ is the total number of corresponding point pair in $S$.
Smaller values of TRE and SRE indicate better registration accuracy.


%
%


In addition to assess registration accuracy using above metrics, some studies also evaluate the quality of the estimated deformation field $\phi$.
The Jacobian matrix $J_{\phi}(p)=\nabla \phi(p)$ evaluates the regularity of the deformation field by capturing the local gradient of $\phi$ around pixel $p$.
Smaller value of the percentage of pixels with non-positive Jacobian determinant (\%$\left | J_{\phi} \right | \leq 0$) indicates smoother $\phi$.

\subsection{PCa Classification and Detection}
Early and accurate diagnosis, along with precise staging, effectively improves the success rate of treatment.
The standard method for diagnosing and grading PCa is the histopathological analysis of prostate tissue samples, typically obtained via TRUS-guided biopsy.
Consequently, the accurate identification of target lesions during the biopsy procedure has been a long-standing and active research task.
To better support TRUS-guided targeted biopsies, automated detection methods have been developed to predict PCa based on TRUS images, see Fig.~\ref{fig:task}(c).
Most of these methods approach cancer detection as a classification problem, where small regions of interest or whole images are categorized as benign or malignant.
We classify and introduce these methods in Section~\ref{subsec:cla}, according to TRUS modality and technical route.

\subsubsection{Evaluation Metrics}
The evaluation metrics for classification task mainly include 
area under the receiver operating characteristic curve (AUC),
accuracy~\eqref{formula:ACC},
sensitivity~\eqref{formula:Sen},
specificity~\eqref{formula:Spe},
precision~\eqref{formula:Precision},
and F-score.

The AUC is a measure of the diagnostic ability of a binary classifier,
which computes the area under the receiver operating characteristic (ROC) curve.
The ROC curve is a curve drawn with the true positive rate (TPR, namely sensitivity) as the Y-axis and the false positive rate (FPR, namely specificity) as the X-axis.
The area under ROC curve can be calculated as:
\begin{equation}
\label{formula:AUC}
\text{AUC} = \int_{0}^{1} \text{TPR}(fpr) \, dfpr.
\end{equation}
It represents the probability that a randomly chosen positive instance is ranked higher than a randomly chosen negative instance.


The F-score is a weighted harmonic mean of precision and recall:
\begin{equation}
\label{formula:F-Score}
\text{F-score} = 2 \times \frac{\text{Precision} \times \text{Recall}}{\text{Precision} + \text{Recall}}.
\end{equation}

\subsection{Needle Detection}
Prostate needles are primarily used during surgical procedures for various purposes, including biopsy, brachytherapy, HIFU, cryotherapy, and so on.
Automated detection and segmentation of interventional needles are crucial in surgical settings because this ensures precise localization and safe interventions, as illustrated in Fig.~\ref{fig:task}(d).
We summarize needle detection methods in Section~\ref{subsec:det}, according to clinical scenario and technical route.

\subsubsection{Evaluation Metrics}
\label{sec:needlemetric}
The numerical metrics for evaluating the needle detection/segmentation mainly include shaft error and tip error.
Shaft error measures the deviation between the predicted and actual axes of the needle, usually expressed in terms of distance or angle:
\begin{equation}
\label{formula:Shaft error distance}
\text{E}_{shaft}^{distance}=\frac{1}{N} \sum_{i=1}^N\left\| \boldsymbol{o}_i, \hat{\boldsymbol{o}_i} \right\|_2,
\end{equation}

\begin{equation}
\label{formula:Shaft error angle}
\text{E}_{shaft}^{angle}=arccos \left(\frac{\boldsymbol{v} \cdot \hat{\boldsymbol{v}}}{\|\boldsymbol{v}\|\|\hat{\boldsymbol{v}}\|}\right),
\end{equation}
where $N$ is the number of points in a needle,
$\hat{\boldsymbol{o}_i}$ represents the predicted position for the i-th point,
and $\boldsymbol{o}_i$ is the ground truth position for the of the i-th point.
$\boldsymbol{v}$ and $\hat{\boldsymbol{v}}$ represent the actual and predicted axis direction vectors.

Tip error measures the difference between the predicted and actual tip positions of the needle:
\begin{equation}
\label{formula:Tip error}
\text{E}_{tip}=\left\| \boldsymbol{t}, \hat{\boldsymbol{t}} \right\|_2,
\end{equation}
where $\boldsymbol{t}$ and $\hat{\boldsymbol{t}}$ represent coordinates of the actual and predicted needle tip.

\section{Method Overview}
\label{sec:method}

In this section, we provide an overview of the key technologies involved in the four tasks of transrectal ultrasound (TRUS) prostate imaging analysis.

\subsection{Prostate Gland Segmentation}
\label{subsec:seg}


\subsubsection{Modalities}
Multi-modality TRUS images including B-mode, color Doppler, elastography, and micro-ultrasound have been utilized for the diagnosis of PCa.
Each modality offers unique insights into the physiological and pathological states of the diseases.
Among the TRUS modalities, the B-mode ultrasound is predominantly employed for prostate segmentation.
The B-mode TRUS affords a morphological representation of the prostate, thereby facilitating the accurate delineation of the prostate's contour.

Almost all segmentation studies mentioned in this survey used B-mode TRUS for the prostate segmentation.
Few studies employed multi-modality TRUS or high-resolution micro-ultrasound.
\cite{seg60} boosted the performance of B-mode TRUS segmentation by offering supplementary information from the vibro-elastography images.
\cite{seg113} and \cite{seg124} segmented prostate gland from the micro-ultrasound images.
Their research has underscored the potential of this high-resolution modality in the precise segmentation of the prostate.



\begin{figure}[t]
	\centering
	\includegraphics[width=\columnwidth]{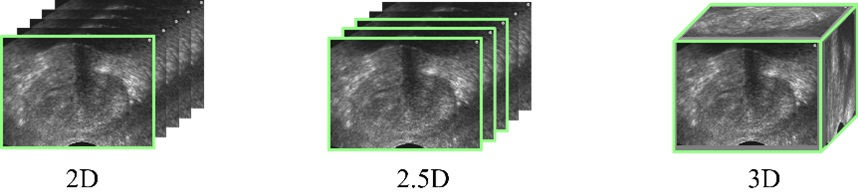}
	\caption{Illustration of different dimensions of the input images for the processing algorithms. The fluorescence indicates the field of view (FOV) handled by the algorithm in a single pass.}
	\label{fig:2.5D}
\end{figure}

\begin{figure*}[t]
	\scriptsize
	\[
	\raisebox{-10ex}{\rotatebox{90}{\textbf{Dimensions (Segmentation)}}}
	\begin{cases}
	\text{2D: }
	\begin{aligned}[t] 
	&\textit{\cite{seg1}}, \textit{\cite{seg2}}, \textit{\cite{seg3}}, \textit{\cite{seg4}}, \textit{\cite{seg6}}, \textit{\cite{seg7}}, \textit{\cite{seg11}},\\
	&\textit{\cite{seg13}}, \textit{\cite{seg15}}, \textit{\cite{seg16}}, \textit{\cite{seg18}}, \textit{\cite{seg19}}, \textit{\cite{seg20}},\\
	&\textit{\cite{seg21}}, \textit{\cite{seg22}}, \textit{\cite{seg23}}, \textit{\cite{seg24}}, \textit{\cite{seg25}}, \textit{\cite{seg27}}, \textit{\cite{seg28}},\\
	&\textit{\cite{seg32}}, \textit{\cite{seg33}}, \textit{\cite{seg34}}, \textit{\cite{seg35}}, \textit{\cite{seg36}}, \textit{\cite{seg37}},\\
	&\textit{\cite{seg38}}, \textit{\cite{seg39}}, \textit{\cite{seg40}}, \textit{\cite{seg41}}, \textit{\cite{seg42}}, \textit{\cite{seg44}}, \textit{\cite{seg45}},\\
	&\textit{\cite{seg46}}, \textit{\cite{seg49}}, \textit{\cite{seg52}}, \textit{\cite{seg53}}, \textit{\cite{seg54}}, \textit{\cite{seg55}},\\
	&\textit{\cite{seg56}}, \textit{\cite{seg57}}, \textit{\cite{seg62}}, \textit{\cite{seg63}}, \textit{\cite{seg69}}, \textit{\cite{seg70}}, \textit{\cite{seg71}},\\
	&\textit{\cite{seg74}}, \textit{\cite{seg75}}, \textit{\cite{seg77}}, \textit{\cite{seg78}}, \textit{\cite{seg79}}, \textit{\cite{seg80}}, \textit{\cite{seg81}},\\
	&\textit{\cite{seg82}}, \textit{\cite{seg85}}, \textit{\cite{seg90}}, \textit{\cite{seg91}}, \textit{\cite{seg92}}, \textit{\cite{seg93}},\\
	&\textit{\cite{seg94}}, \textit{\cite{seg96}}, \textit{\cite{seg99}}, \textit{\cite{seg100}}, \textit{\cite{seg101}}, \textit{\cite{seg102}}, \textit{\cite{seg103}},\\
	&\textit{\cite{seg104}}, \textit{\cite{seg105}}, \textit{\cite{seg106}}, \textit{\cite{seg107}}, \textit{\cite{seg108}}, \textit{\cite{seg111}}, \textit{\cite{seg112}},\\
	&\textit{\cite{seg113}}, \textit{\cite{seg114}}, \textit{\cite{seg115}}, \textit{\cite{seg116}}, \textit{\cite{seg117}}, \textit{\cite{seg118}}, \textit{\cite{seg121}},\\
	&\textit{\cite{seg122}}, \textit{\cite{seg123}}\\
	\end{aligned}\\
	\\
	\text{2.5D: }
	\textit{\cite{seg86}}, \textit{\cite{seg87}}, \textit{\cite{seg97}}\\
	\\
	\text{3D: }
	\begin{aligned}[t] 
	&\textit{\cite{seg8}}, \textit{\cite{seg9}}, \textit{\cite{seg10}}, \textit{\cite{seg12}}, \textit{\cite{seg14}}, \textit{\cite{seg17}}, \textit{\cite{seg26}},\\
	&\textit{\cite{seg29}}, \textit{\cite{seg30}}, \textit{\cite{seg31}}, \textit{\cite{seg38}}, \textit{\cite{seg43}}, \textit{\cite{seg47}},\\
	&\textit{\cite{seg48}}, \textit{\cite{seg51}}, \textit{\cite{seg58}}, \textit{\cite{seg59}}, \textit{\cite{seg60}}, \textit{\cite{seg61}}, \textit{\cite{seg64}},\\
	&\textit{\cite{seg65}}, \textit{\cite{seg66}}, \textit{\cite{seg67}}, \textit{\cite{seg68}}, \textit{\cite{seg72}}, \textit{\cite{seg73}}, \textit{\cite{seg76}},\\
	&\textit{\cite{seg83}}, \textit{\cite{seg84}}, \textit{\cite{seg88}}, \textit{\cite{seg89}}, \textit{\cite{seg95}}, \textit{\cite{seg98}}, \textit{\cite{seg109}},\\
	&\textit{\cite{seg110}}, \textit{\cite{seg119}}, \textit{\cite{seg124}}\\
	\end{aligned}\\
	\end{cases}
	\]
	\caption{Prostate segmentation approaches categorized as 2D, 2.5D, and 3D methods.}
	\label{fig:seg_dimension}
\end{figure*}

\subsubsection{Dimensions}
We categorize prostate segmentation approaches as 2D, 2.5D, 3D methods, according to the TRUS dimension they processed.
Note that 2D and 3D algorithms process one slice and a whole volume, respectively, while 2.5D processes several consecutive 2D slices (see Fig.~\ref{fig:2.5D}).
As depicted in Fig.~\ref{fig:seg_dimension}, the majority of researches have concentrated on 2D and 3D techniques, with a comparatively modest exploration of the 2.5D methodology.

The 2D segmentation is executed on an individual slice of the prostate image. 
This approach is frequently employed for intraoperative real-time navigation, serving to assist clinicians in the precise identification of the prostate gland's location and boundary.
The merits of 2D segmentation include its simplicity and high computational efficiency.
Nonetheless, it falls short in capturing the 3D morphology of the prostate, potentially leading to inaccuracies in the volumetric assessment of the prostate, particularly concerning the apex and base regions.

In contrast to 2D segmentation, three-dimensional segmentation is conducted on the entire volumetric TRUS data.
This technique is predominantly utilized in preoperative planning, aiding physicians in delineating the extent and trajectory for surgical excision or in radiation therapy planning by enabling radiation oncologists to pinpoint the target region for irradiation.
3D segmentation affords an exhaustive and accurate quantitative analysis of the prostate's shape, volume, and boundaries.
However, it is encumbered by high computational demands and necessitates the processing of a vast array of voxels and intricate 3D structures.

Beyond the 2D and 3D approaches, the 2.5D segmentation technique operates on multiple contiguous slices of the prostate image.
By integrating information from multiple slices, 2.5D segmentation can partially capture the volumetric information.
It offers an advantage over 2D segmentation in better handling the 3D morphological variations of the prostate, albeit with limitations imposed by the selection and thickness of the slices.
Compared with 3D segmentation, 2.5D segmentation can alleviate the computational load, presenting a more efficient compromise between computational demands and segmentation accuracy.

\begin{figure*}[t]
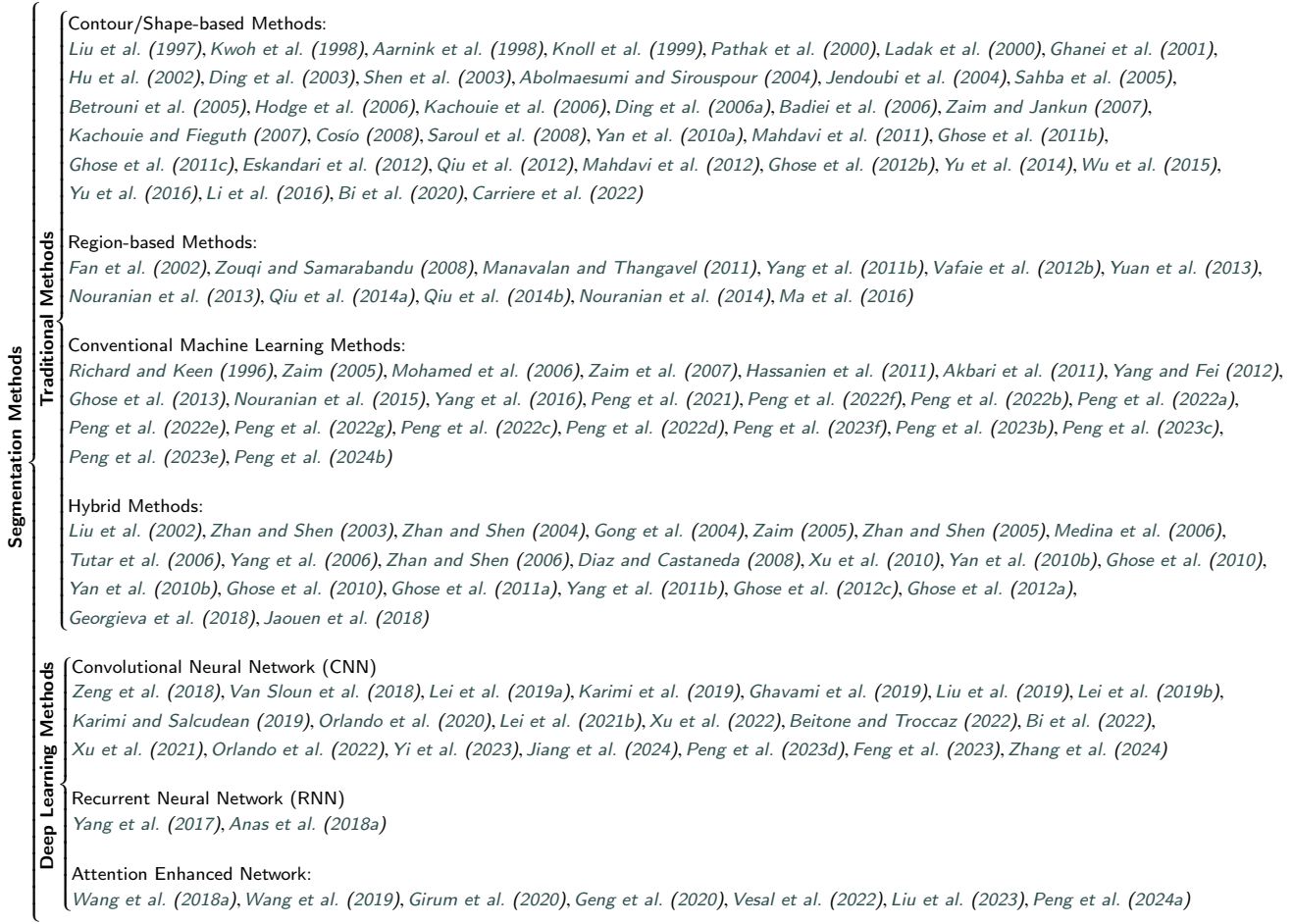

	\scriptsize
	\[
	\raisebox{-10ex}{\rotatebox{90}{\textbf{Segmentation Methods}}}
	\begin{cases}
	\raisebox{-10ex}{\rotatebox{90}{\textbf{Traditional Methods}}}
	\begin{cases}
	\text{Contour/Shape-based Methods:}\\
	\begin{aligned}[t]
	&\textit{\cite{seg2}}, \textit{\cite{seg3}}, \textit{\cite{seg4}}, \textit{\cite{seg5}}, \textit{\cite{seg6}}, \textit{\cite{seg7}}, \textit{\cite{seg8}},\\
	&\textit{\cite{seg9}},\textit{\cite{seg12}},\textit{\cite{seg13}},\textit{\cite{seg15}},\textit{\cite{seg16}},\textit{\cite{seg19}},\\
	&\textit{\cite{seg20}},\textit{\cite{seg24}},\textit{\cite{seg25}},\textit{\cite{seg26}},\textit{\cite{seg27}},\textit{\cite{seg32}},\\
	&\textit{\cite{seg33}},\textit{\cite{seg35}},\textit{\cite{seg36}},\textit{\cite{seg41}},\textit{\cite{seg43}},\textit{\cite{seg46}},\\
	&\textit{\cite{seg49}},\textit{\cite{seg57}},\textit{\cite{seg59}},\textit{\cite{seg60}},\textit{\cite{seg53}},\textit{\cite{seg65}},\textit{\cite{seg69}},\\
	&\textit{\cite{seg71}},\textit{\cite{seg74}},\textit{\cite{seg93}},\textit{\cite{seg106}}
	\end{aligned}
	\\ \\
	\text{Region-based Methods:}\\
	\begin{aligned}[t]
	&\textit{\cite{seg10}},\textit{\cite{seg37}},\textit{\cite{seg52}},\textit{\cite{seg48}},\textit{\cite{seg62}},\textit{\cite{seg61}},\\
	&\textit{\cite{seg64}},\textit{\cite{seg66}},\textit{\cite{seg67}},\textit{\cite{seg68}},\textit{\cite{seg73}}
	\end{aligned}
	\\ \\
	\text{Conventional Machine Learning Methods:}\\
	\begin{aligned}[t]
	&\textit{\cite{seg1}},\textit{\cite{seg21}},\textit{\cite{seg28}},\textit{\cite{seg34}},\textit{\cite{seg45}},\textit{\cite{seg47}},\textit{\cite{seg58}},\\
	&\textit{\cite{seg63}},\textit{\cite{seg70}},\textit{\cite{seg72}},\textit{\cite{seg96}},\textit{\cite{seg102}},\textit{\cite{seg103}},\textit{\cite{seg104}},\\
	&\textit{\cite{seg105}},\textit{\cite{seg107}},\textit{\cite{seg108}},\textit{\cite{seg101}},\textit{\cite{seg114}},\textit{\cite{seg115}},\textit{\cite{seg117}},\\
	&\textit{\cite{seg121}},\textit{\cite{seg123}}
	\end{aligned}
	\\ \\
	\text{Hybrid Methods:}\\
	\begin{aligned}[t]
	&\textit{\cite{seg11}},\textit{\cite{seg14}},\textit{\cite{seg17}},\textit{\cite{seg18}},\textit{\cite{seg21}},\textit{\cite{seg22}},\textit{\cite{seg23}},\\
	&\textit{\cite{seg29}},\textit{\cite{seg30}},\textit{\cite{seg31}},\textit{\cite{seg38}},\textit{\cite{seg42}},\textit{\cite{seg39}},\textit{\cite{seg40}},\\
	&\textit{\cite{seg39}},\textit{\cite{seg40}},\textit{\cite{seg44}},\textit{\cite{seg48}},\textit{\cite{seg54}},\textit{\cite{seg55}},\\
	&\textit{\cite{seg80}},\textit{\cite{seg89}}
	\end{aligned}\\
	\end{cases}
	\\ \\
	\raisebox{-10ex}{\rotatebox{90}{\textbf{Deep Learning Methods}}}
	\begin{cases}
	\text{Convolutional Neural Network (CNN)}\\
	\begin{aligned}[t]
	&\textit{\cite{seg79}},\textit{\cite{seg82}},\textit{\cite{seg84}},\textit{\cite{seg85}},\textit{\cite{seg86}},\textit{\cite{seg87}},\textit{\cite{seg88}},\\
	&\textit{\cite{seg91}},\textit{\cite{seg90}},\textit{\cite{seg95}},\textit{\cite{seg98}},\textit{\cite{seg99}},\textit{\cite{seg100}},\\
	&\textit{\cite{seg109}},\textit{\cite{seg110}},\textit{\cite{seg111}},\textit{\cite{seg113}},\textit{\cite{seg116}},\textit{\cite{seg118}},\textit{\cite{seg124}}\\
	\end{aligned}
	\\ \\
	\text{Recurrent Neural Network (RNN)}\\
	\textit{\cite{seg75}},\textit{\cite{seg78}}
	\\ \\
	\text{Attention Enhanced Network:}\\
	\begin{aligned}[t]
	&\textit{\cite{seg77}},\textit{\cite{seg83}},\textit{\cite{seg92}},\textit{\cite{seg94}},\textit{\cite{seg97}},\textit{\cite{seg119}},\textit{\cite{seg122}}
	\end{aligned} 
	\end{cases}
	\end{cases}
	\]
	\caption{Traditional segmentation methods and deep learning (DL) methods. Traditional methods can be categorized as contour/shape based methods, region based methods, conventional machine learning methods, and hybrid methods. DL methods mainly apply convolutional neural network (CNN), recurrent neural network (RNN), and attention enhanced networks.}
	\label{fig:seg_method}
\end{figure*}

\subsubsection{Segmentation Methods}
We categorize prostate segmentation methods into two ways:
traditional methods and deep learning (DL) methods.
Fig.~\ref{fig:seg_method} shows the categorization in detail.
It can be seen that before 2016 almost all of the studies were about traditional methods, and after that DL methods dominated this area.

\begin{table*}[t]
	\centering
	\caption{Overview of traditional methods for prostate gland segmentation in TRUS.}
	\fontsize{5}{10}\selectfont
	\begin{tabular}{cccccccc}
		\toprule
		\multirow{2}{*}{\parbox{0.8cm}{\centering Traditional\\Methods}}   &\multirow{2}{*}{\centering References} &\multirow{2}{*}{\centering Dimension}  &\multirow{2}{*}{\centering Auto/Semi-auto} & \multirow{2}{*}{\centering Methods} & \multirow{2}{*}{\centering Samples} & \multicolumn{2}{c}{Results} \\
		\cmidrule{7-8} 
		& & & & & & Metrics & Values \\
		\midrule
		\multirow{14}{*}{\parbox{0.8cm}{\centering Contour\\and\\Shape}}
		
		& \cite{seg6} & 2D & Semi-auto & Probabilistic filters, edge detection, manual editing & 125 & \parbox{2cm}{\centering MAD\\HD} & \parbox{2cm}{\centering \setstretch{0.8} \fontsize{6}{12} $0.7\pm0.4$ $\text{mm}$ \\ $1.8\pm1.0$  $\text{mm}$}\\
		\cmidrule{7-8}
		
		& \cite{seg13} & 2D & Automatic & Gabor filter bank, statistical shape model & 8 & \parbox{2cm}{\centering 1-DSC\\1-IoU\\MAD}    
		& \parbox{2cm}{\centering \setstretch{0.8} \fontsize{6}{12} $1.66\pm1.68\%$ \\ $3.98\pm0.97\%$ \\ $3.20\pm0.87$ $\text{pixels}$}\\
		\cmidrule{7-8}
		
		
		& \cite{seg27} & 2D & Semi-auto & Image warping, edge detection, ellipse fitting & 17 & \parbox{2cm}{\centering MAD\\MAXD} & \parbox{2cm}{\centering \setstretch{0.8} \fontsize{6}{12}$0.67\pm0.18$ $\text{mm}$ \\ $2.25\pm0.56$ $\text{mm}$}\\
		\cmidrule{7-8}
		
		& \cite{seg35} & 2D & Automatic & Pixel classification, active shape model & 22 & \parbox{2cm}{\centering MAD\\MAXD} & \parbox{2cm}{\centering \setstretch{0.8} \fontsize{6}{12} $1.74\pm0.69$ $\text{mm}$ \\ $4.51\pm2.42$ $\text{mm}$}\\
		\cmidrule{7-8}
		
		
		& \cite{seg43} & 3D & Semi-auto & Shape fitting, convex optimization & 40 & \parbox{2cm}{\centering 1-DSC} & \parbox{2cm}{\centering \setstretch{0.8} \fontsize{6}{12} $5.82\pm4.15\%$}\\
		\cmidrule{7-8}
		
		
		& \cite{seg59} & 3D & Semi-auto & Shape constraint, level set & 20 & \parbox{2cm}{\centering DSC\\MAD}     
		&\parbox{2cm}{\centering \setstretch{0.8} \fontsize{6}{12}$93.39\pm1.26\%$\\ $1.16\pm0.34$ $\text{mm}$}\\
		\cmidrule{7-8}
		
		& \cite{seg60} & 3D & Automatic & Elastography, active shape model & 11 &\parbox{2cm}{\centering DSC } & \parbox{2cm}{\centering \setstretch{0.8} \fontsize{6}{12} $87\pm7\%$ }\\
		\cmidrule{7-8}
		
		& \cite{seg69} & 2D & Automatic & Rotation-invariant texture feature, graph cut, level set & 598 & \parbox{2cm}{\centering MAD} 
		& \parbox{2cm}{\centering \setstretch{0.8} \fontsize{6}{12} $1.06\pm0.53$ $\text{mm}$}\\
		\cmidrule{7-8}
		
		& \cite{seg93} & 2D & Automatic & Rayleigh mixture model, active shape model & 30 & \parbox{2cm}{\centering DSC\\MAD} & \parbox{2cm}{\centering \setstretch{0.8} \fontsize{6}{12} $95\pm0.81\%$ \\ $1.86\pm0.02$ $\text{pixels}$}\\
		\midrule
		
		\multirow{14}{*}{\centering  Region}
		
		& \cite{seg10} & 3D & Semi-auto & Region information, level set & 8 & Visual Comparison & Visual Comparison\\
		\cmidrule{7-8}
		
		
		& \cite{seg61} & 3D & Automatic & Rotational symmetry, convex optimization & 20 & \parbox{2cm}{\centering DSC\\MAD\\MAXD} & \parbox{2cm}{\centering \setstretch{0.8} \fontsize{6}{12} $93.7\pm2.1\%$ \\ $1.12\pm0.4$ $\text{mm}$ \\ $3.15\pm0.65$ $\text{mm}$} \\
		\cmidrule{7-8}
		
		& \cite{seg64} & 3D & Automatic & Multi-atlas fusion, pairwise shape similarity & 50 & \parbox{2cm}{\centering DSC\\MAD\\MAXD} & \parbox{2cm}{\centering \setstretch{0.8} \fontsize{6}{12} $92.08\pm1.96\%$ \\ $1.22\pm0.57$ $\text{mm}$ \\ $3.47\pm1.48$ $\text{mm}$}\\
		\cmidrule{7-8}
		
		& \cite{seg66} & 3D & Automatic & Volume-preserving prior, convex optimization & 12 & \parbox{2cm}{\centering DSC\\MAD\\MAXD} & \parbox{2cm}{\centering \setstretch{0.8} \fontsize{6}{12} $89.5\pm2.4\%$ \\ $1.4\pm0.6$ $\text{mm}$ \\ $5.2\pm3.2$ $\text{mm}$}\\
		\cmidrule{7-8}
		
		& \cite{seg67} & 3D & Automatic & Axial symmetry, convex optimization & 25 & \parbox{2cm}{\centering DSC} & \parbox{2cm}{\centering \setstretch{0.8} \fontsize{6}{12}$93.2\pm2.0\%$}\\
		\cmidrule{7-8}
		
		& \cite{seg68} & 3D & Automatic & Multi-atlas fusion, atlas agreement factor & 280 & \parbox{2cm}{\centering DSC\\MSD} & \parbox{2cm}{\centering \setstretch{0.8} \fontsize{6}{12} $89.75\pm3.75\%$ \\ $1.33\pm0.60$ $\text{mm}$}\\
		\midrule
		
		\multirow{11}{*}{\parbox{0.8cm}{\centering Machine\\Learning}} 
		
		& \cite{seg63} & 2D & Automatic & Statistical shape and appearance model, random forest & 126 & \parbox{2cm}{\centering DSC\\MAD\\HD} & \parbox{2cm}{\centering \setstretch{0.8} \fontsize{6}{12} $91\pm4\%$ \\$1.30\pm0.3$ $\text{mm}$ \\$3.62\pm0.56$ $\text{mm}$}\\
		\cmidrule{7-8}
		
		& \cite{seg70} & 2D & Automatic & Sparse dictionary learning & 590 & \parbox{2cm}{\centering MSD\\HD} & \parbox{2cm}{\centering \setstretch{0.8} \fontsize{6}{12} $0.98\pm0.39$ $\text{mm}$ \\ $5.4\pm1.38$ $\text{mm}$}\\
		\cmidrule{7-8}
		
		& \cite{seg72} & 3D & Automatic & Kernel support vector machine & 10 & \parbox{2cm}{\centering DSC} & \parbox{2cm}{\centering \setstretch{0.8} \fontsize{6}{12} $89.7\pm2.3\%$}\\
		\cmidrule{7-8}
		
		
		& \cite{seg103} & 2D & Semi-auto & Improved closed principal curve & 50 & \parbox{2cm}{\centering DSC\\Jaccard\\ACC} & \parbox{2cm}{\centering \setstretch{0.8} \fontsize{6}{12}$96.5\%$\\$95.1\%$\\$96.3\%$}\\
		\cmidrule{7-8}
		
		
		
		& \cite{seg123}  & 2D & Semi-auto & Adaptive polygon tracking model, principal curve & 1320 & \parbox{2cm}{\centering DSC\\Jaccard\\ACC} & \parbox{2cm}{\centering \setstretch{0.8} \fontsize{6}{12} $96.4\pm2.4\%$ \\ $95.5\pm2.9\%$ \\ $96.1\pm2.5\%$}\\
		\midrule
		
		\multirow{13}{*}{\centering  Hybrid} 
		
		& \cite{seg14} & 3D & Automatic & Kernel support vector machine, statistical shape model & 3 & \parbox{2cm}{\centering 1-IoU\\MAD} & \parbox{2cm}{\centering \setstretch{0.8} \fontsize{6}{12} $3.9\%$ \\ $0.81$ $\text{voxel}$}\\
		\cmidrule{7-8}
		
		& \cite{seg18} & 2D & Automatic & Deformable superellipses, Bayesian model & 125 & \parbox{2cm}{\centering MAD\\HD} & \parbox{2cm}{\centering \setstretch{0.8} \fontsize{6}{12} $1.36\pm0.58$ $\text{mm}$\\$3.42\pm1.52$ $\text{mm}$}\\
		\cmidrule{7-8}
		
		& \cite{seg29} & 3D & Semi-auto & Spherical harmonics, statistical analysis & 30 & \parbox{2cm}{\centering MAD\\MAXD} & \parbox{2cm}{\centering \setstretch{0.8} \fontsize{6}{12} $1.26\pm0.41$ $\text{mm}$\\ $4.06\pm1.25$ $\text{mm}$}\\
		\cmidrule{7-8}
		
		& \cite{seg31} & 3D & Automatic & Gabor-support vector machine, statistical texture matching & 5 & \parbox{2cm}{\centering 1-IoU\\MAD} & \parbox{2cm}{\centering \setstretch{0.8} \fontsize{6}{12} $4.31\pm0.40\%$ \\ $1.07\pm0.10$ $\text{voxels}$}\\
		\cmidrule{7-8}
		
		& \cite{seg39} & 2D & Automatic & Partial active shape model, region model & 301 & \parbox{2cm}{\centering MAD} & \parbox{2cm}{\centering \setstretch{0.8} \fontsize{6}{12}$2.01\pm1.02$ $\text{mm}$}\\
		\cmidrule{7-8}
		
		& \cite{seg51} & 3D & Semi-auto & Discrete dynamic contour, optimal surface detection & 28 & \parbox{2cm}{\centering ASD} & \parbox{2cm}{\centering \setstretch{0.8} \fontsize{6}{12}$0.77\pm0.28$ $\text{mm}$}\\
		\cmidrule{7-8}
		
		
		& \cite{seg89} & 3D & Automatic & Hybrid edge/region-based parametric deformable model & 36 & \parbox{2cm}{\centering DSC\\MAD\\HD} & \parbox{2cm}{\centering \setstretch{0.8} \fontsize{6}{12} $92\pm2\%$ \\ $1.42\pm0.40$ $\text{mm}$\\ $4.03\pm1.22$ $\text{mm}$}\\
		\bottomrule
	\end{tabular}
	\label{table:seg_trad}
\end{table*}

\paragraph{\textbf{Traditional Methods.}}
We present the traditional methods for prostate segmentation in four categories:
contour/shape-based methods, region-based methods, conventional machine learning methods, and hybrid methods.
Such categorization is determined on the underlying theoretical computational approaches employed to address the segmentation task.
Table~\ref{table:seg_trad} gives a comprehensive summary of the representative studies using traditional methods for TRUS segmentation.

\textbf{(a) Contour/Shape-based Methods.}
These methods use prostate contour/edge information to guide the segmentation.
Contour-based methods typically initiate with edge detection.
\cite{seg6} introduced a Sticks algorithm to enhance contrast and reduce speckle noise in TRUS, thus facilitating the rendering of prostate edges as a visual guide for subsequent edge detection and manual refinement.
\cite{seg13} expanded on this by employing a Gabor filter bank to characterize prostate boundary features across multiple scales and orientations, followed by a hierarchical deformation strategy to adaptively focus on feature similarity. 
This approach, while more time-consuming, captured a broader range of boundary directions.
\cite{seg69} diverged from the traditional view by leveraging speckles' intrinsic properties to aid segmentation, arguing that speckle characteristics provide useful cues for prostate boundary delineation.

Active shape models (ASM), as pioneered by~\cite{cootes1995active}, offer a statistical shape-based method that has been further developed by researchers such as~\cite{seg35}, who presented a robust automatic initialization of the ASM for prostate.
This method used pixel classification for the initialization of prostate region and a multi-population genetic algorithm for ASM pose adjustment. 
\cite{seg60} and \cite{seg93} extended the ASM by incorporating high-contrast vibro-elastography images and clustering techniques using a Rayleigh mixture model, respectively, to enhance the initialization and deformation guidance of the ASM model.

Beyond ASM, other shape-prior-based segmentation techniques have been explored.
\cite{seg27} proposed an image warping technique to impose an elliptical shape on the prostate, with subsequent optimization of the elliptical fitting using boundary points from an edge detector.
\cite{seg43} further advanced this by suggesting twisted and tapered ellipsoids as suitable a priori 3D shapes, transforming the shape fitting into a convex problem for efficient solution.
\cite{seg41} proposed adaptive learning of local neighborhood shape statistics to capture patient-specific variations.
\cite{seg59} introduced a semi-automatic method integrating shape constraints with a modified level set formulation for 3D segmentation.
These methods emphasized the importance of integrating shape priors with computational strategies to improve accuracy and efficiency.

\textbf{(b) Region-based Methods.}
These methods harness local intensity values or statistical measures within an energy minimization framework to achieve segmentation.
\cite{seg10} initially developed a fast discrimination technique for extracting the prostate region and subsequently addressed the ``boundary leaking'' issue by integrating regional information, rather than gradient information, into the level set method.
\cite{seg61} employed rotationally re-sliced images around a specified axis to delineate 3D prostate boundaries, a technique that effectively capitalized on the inherent rotational symmetry of prostate shapes to adjust a series of 2D slice segmentation in a unified 3D context.
The work~\citep{seg61} was further extended to~\citep{seg67}, with more validation data.
\cite{seg66} introduced a convex optimization strategy for extracting the prostate surface, while maintaining the global volume size prior.
Moreover, they implemented an efficient numerical solver on GPUs.
Additionally, some methods accomplish the extraction of the prostate region through template matching.
\cite{seg64} designed a multi-atlas fusion framework to segment TRUS images using pairwise atlas shape similarity.
\cite{seg68} further introduced the atlas agreement factor to boost the robustness of the previous work~\citep{seg64}.
This factor was integrated into an atlas selection algorithm to refine the dataset prior to merging the atlas contours, yielding consistent segmentation.

\textbf{(c) Conventional Machine Learning Methods.}
Early prostate segmentation methods also encompass machine learning approaches.
These techniques leverage features such as intensity or higher-dimensional attributes like filter responses to cluster and/or classify the image into prostate and background regions.
\cite{seg63} introduced a method for prostate segmentation that constructed a multiple average parameter model derived from the principal component analysis of shape and a posterior probabilities within a multi-resolution framework.
The model parameters were subsequently refined using a priori knowledge of the optimization space to achieve optimal prostate segmentation.
\cite{seg70} proposed a learning-based multi-label segmentation algorithm.
A sparse representation method was integrated into the algorithm to learn a dictionary encoding information of images as well as clinical and planning target volume segmentation.
\cite{seg72} developed a block-based feature learning framework, where patient-specific anatomical features extracted from aligned training images served as signatures for each voxel.
A feature selection process identified the most robust and informative features to train a kernel support vector machine (KSVM).

Additionally, some methods have employed the principal curve, which provided a means of representing low-dimensional data manifolds in high-dimensional spaces, thereby effectively capturing the complexity of prostate shapes. 
\cite{seg103} proposed an improved closed principal curve approach, utilizing only a few seed points defined by radiologists as an approximate initialization.
\cite{seg123} further developed an adaptive polygon tracking model to alleviate the issue of determining the segment number of the principal curve automatically.
In addition, an interpretable mathematical function was formulated to constrain a smooth boundary.

\textbf{(d) Hybrid Methods.}
The objective of hybrid methods is to leverage advantages from contour, shape, region, and conventional machine learning to achieve better segmentation results.
\cite{seg14} combined kernel support vector machine and statistical shape model (SSM) to segment 3D TRUS.
The KSVM was employed to classify the prostate from surrounding tissues, and then the SSM was used to refine the contour.
Later, \cite{seg31} improved~\citep{seg14} by using a set of Gabor-support vector machines (G-SVMs) and the statistical texture matching model.
\cite{seg18} emphasized that the prior knowledge of prostate shapes played an important role for reliable segmentation.
They used deformable superellipses to model the shape of prostate and further developed a robust Bayesian segmentation approach.
\cite{seg29} employed the eighth spherical harmonic function to depict the shape of the prostate, and statistically analyzed the shape parameters.
\cite{seg39} employed partial active shape model and region model to address the issue of missing boundaries in shadow areas.
\cite{seg51} sequentially combined discrete dynamic contour and optimal surface detection for semi-automatic 3D TRUS segmentation in HIFU Therapy.
\cite{seg89} proposed a hybrid edge and region-based parametric deformable model.
The edge-based model was derived from the radial bas-relief approach, while the region-based model was regarded as a parametric active surface driven by the Bhattacharyya gradient flow.

\paragraph{\textbf{Deep Learning Methods.}}
We categorize deep learning methods for TRUS segmentation into three classes:
convolutional neural network (CNN), recurrent neural network (RNN), and attention enhanced networks, according to their network structure.
Due to limited studies on Transformer currently in this field, relevant researches are included in the attention enhanced networks section.
Table~\ref{table:seg_dl} summarizes representative studies related to the DL-based segmentation in TRUS.

\begin{table*}[t]
	\centering
	\caption{Overview of deep learning methods for prostate gland segmentation in TRUS.}
	\fontsize{5}{10}\selectfont
	\begin{tabular}{ccccccc}
		\toprule
		\multirow{2}{*}{\parbox{1.2cm}{\centering Deep Learning\\Methods}}  &\multirow{2}{*}{\centering References} &\multirow{2}{*}{\centering Dimension}  & \multirow{2}{*}{\centering Methods} & \multirow{2}{*}{\centering Samples} & \multicolumn{2}{c}{Results} \\
		\cmidrule{6-7} 
		& & & & & Metrics & Values \\
		\midrule
		\multirow{16}{*}{\centering CNN}
		
		& \cite{seg84} & 3D & \parbox{5cm}{ 1) Multi-directional deeply supervised V-Net \\ 2) A hybrid loss with binary cross-entropy and batch-based Dice} & 44 & \parbox{2cm}{\centering DSC\\MSD\\HD} & \parbox{2cm}{\centering $92\pm3\%$ \\ $0.60\pm0.23$ $\text{mm}$ \\ $3.94\pm1.55$ $\text{mm}$ }\\
		\cmidrule{6-7}
		
		& \cite{seg85} & 2D & \parbox{5cm}{ 1) An adaptive sampling strategy to address difficult cases \\ 2) A CNN ensemble to identify uncertain segmentation} & 675
		& \parbox{2cm}{\centering DSC\\HD} & \parbox{2cm}{\centering $93.9\pm3.5\%$ \\$2.7\pm2.3$ $\text{mm}$}\\
		\cmidrule{6-7}
		
		
		& \cite{seg91} & 2D & \parbox{5cm}{ 1) A loss function to reduce the Hausdorff distance (HD) \\ 2) Three methods to estimate HD} & 2625 & \parbox{2cm}{\centering DSC\\ASD\\HD} & \parbox{2cm}{\centering $94.6\pm4.1\%$ \\ $1.05\pm0.46$ $\text{mm}$ \\ $2.6\pm1.8$ $\text{mm}$ }\\
		\cmidrule{6-7}
		
		& \cite{seg95} & 3D & \parbox{5cm}{1) An anchor-free mask CNN} & 83 & \parbox{2cm}{\centering DSC\\95HD} & \parbox{2cm}{\centering $94\pm3\%$ \\ $2.27\pm0.79$ $\text{mm}$}\\
		\cmidrule{6-7}
		
		& \cite{seg98} & 3D & \parbox{5cm}{ 1) Represent prostate shape in the polar coordinate space \\ 2) A centroid perturbed test-time augmentation strategy} & 315 & \parbox{2cm}{\centering DSC\\ASD\\HD} & \parbox{2cm}{\centering $89.97\pm3.47\%$ \\ $ 1.30\pm0.61$ $\text{mm}$ \\ $7.07\pm3.19$ $\text{mm}$}\\
		\cmidrule{6-7}
		
		& \cite{seg109} & 3D & \parbox{5cm}{ 1) Shadow augmentation and shadow dropout \\ 2) Semi-supervised Learning} & 2423 & \parbox{2cm}{\centering DSC\\ASD\\HD} & \parbox{2cm}{\centering $91.60\pm2.37\%$ \\ $1.02\pm0.34$ $\text{mm}$ \\ $6.37\pm2.36$ $\text{mm}$}\\
		\cmidrule{6-7}
		
		& \cite{seg110} & 2D & \parbox{5cm}{ 1) Modified U-Net and U-Net$++$} & 6721 & \parbox{2cm}{\centering DSC\\MSD\\HD} & \parbox{2cm}{\centering $94.0\%$ \\ $0.90$ $\text{mm}$ \\ $3.27$ $\text{mm}$}\\
		
		
		\midrule
		
		\multirow{3}{*}{\centering RNN} 
		
		& \cite{seg75} & 2D & \parbox{5cm}{ 1) Serialize static TRUS images into dynamic sequences \\ 2) A multi-view shape fusion strategy \\ 3) A multiscale auto-context scheme}  & 530 & \parbox{2cm}{\centering DSC\\Recall\\Precision} & \parbox{2cm}{\centering $92.33\%$ \\ $89.76\%$ \\ $95.19\%$}\\
		\cmidrule{6-7}
		
		& \cite{seg78} & 2D & \parbox{5cm}{ 1) Residual convolution in recurrent networks \\ 2) Recurrent connections for deep network layers}  & 3892 & \parbox{2cm}{\centering DSC\\HD} & \parbox{2cm}{\centering $92.91\pm2.74\%$ \\ $2.97\pm1.96$ $\text{mm}$}\\
		\midrule
		
		\multirow{10}{*}{\centering Attention}
		
		& \cite{seg77} & 2D & \parbox{5cm}{ 1) Select useful complementary information from multi-level features to refine the features at each individual layer } & 530 & \parbox{2cm}{\centering DSC\\Recall\\Precision} & \parbox{2cm}{\centering  $95.27\%$ \\$96.98\%$ \\$93.69\%$}\\
		\cmidrule{6-7}
		
		& \cite{seg83} & 3D & \parbox{5cm}{ 1) Selectively leverage multi-level features from different layers} & 40 & \parbox{2cm}{\centering DSC\\Recall\\Precision } & \parbox{2cm}{\centering  $90\pm3\%$\\$91\pm4\%$\\$90\pm6\%$}\\
		\cmidrule{6-7}
		
		& \cite{seg92} & 2D & \parbox{5cm}{1) A channel-wise attention strategy \\ 2) Learning-based prior shape knowledge} & 145 & \parbox{2cm}{\centering DSC\\95HD} & \parbox{2cm}{\centering $88\pm3\%$\\ $2.01\pm0.54$ $\text{mm}$}\\
		\cmidrule{6-7}
		
		& \cite{seg97} & 2.5D & \parbox{5cm}{ 1) Feature positioning information \\ 2) A training scheme that utilizes knowledge distillation} & 2892 & \parbox{2cm}{\centering DSC\\95HD} & \parbox{2cm}{\centering  $94.0\pm3.0\%$ \\ $2.29\pm1.45$ $\text{mm}$}\\
		\cmidrule{6-7}
		
		& \cite{seg122} & 2D & \parbox{5cm}{ 1) Attention gate modules with squeeze-and-excitation U-net \\ 2) An improved polygon searching model \\ 3) A storage-based quantum evolution model} & 945 & \parbox{2cm}{\centering DSC\\ACC} & \parbox{2cm}{\centering  $94.3\pm3.8\%$ \\ $93.9\pm4.1\%$ }\\
		
		\bottomrule
	\end{tabular}
	\label{table:seg_dl}
\end{table*}

\textbf{(a) Convolutional Neural Network.}
CNNs are among the most successful and widely used architectures in DL, particularly for computer vision tasks.
Following the advent of U-net~\cite{seg125}, a plethora of CNNs with encoder-decoder structures have been employed for medical image segmentation.
Several methods have sought to enhance segmentation performance by improving loss functions.
For instance, \cite{seg84} developed a multi-directional DL method to segment TRUS for US-guided radiation therapy.
They combined binary cross-entropy (BCE) loss and batch-based Dice loss into a staged hybrid loss function for deeply-supervised training.
\cite{seg85} improved the uncertain segmentation by incorporating prior shape information in the form of a statistical shape model.
\cite{seg91} presented a loss function for training the segmentation network with the goal of directly reducing Hausdorff distance.

Additionally, other methods have focused on the improvement of the network structures. 
\cite{seg95} developed an anchor-free mask CNN that consisted of three subnetworks: a backbone, a fully convolutional one-state object detector, and a mask head.
\cite{seg98} proposed a novel polar transform network to address the issue of difficult segmentation in the challenging base and apex regions.
The proposed network represented the prostate shape in the polar coordinate space.
This representation provided much denser samples of the prostate volume, thus facilitating the learning of discriminative features for accurate prostate segmentation.
\cite{wang2024lightcm} combined CNNs and tokenized multilayer perceptron (MLPs) with a pyramid structure and hybrid attention to enable real-time TRUS prostate segmentation, enhancing accuracy through boundary-constrained loss.

Furthermore, some studies have focused on fully utilizing data information.
\cite{seg109} proposed a shadow-consistent semi-supervised segmentation method that included shadow augmentation (Shadow-AUG) and shadow dropout (Shadow-DROP).
Shadow-AUG enlarged training samples by including simulated shadow artifacts to the images, while Shadow-DROP enhanced the network to leverage neighboring high-quality pixels to infer boundaries.
\cite{seg110} developed an automatic 3D TRUS segmentation network, trained on a large, clinically diverse dataset with variable image quality.
\cite{li2024bi} used a CNN network for automatic localization, introduced semantic constraints to reduce noise, and employed bi-directional iteration with denoising for optimal prostate segmentation.

\textbf{(b) Recurrent Neural Network.}
RNNs are commonly utilized for processing sequential data, such as speech, text, videos, and time-series data.
Unlike unidirectional feedforward neural networks, RNNs are bidirectional neural networks, allowing the outputs of certain nodes to influence subsequent inputs from the same nodes.
Their capability to process arbitrary input sequences using internal state, or memory, makes them suitable for tasks such as unsegmented, connected handwriting recognition or speech recognition.

In the context of TRUS segmentation, \cite{seg75} first serialized static 2D TRUS images into dynamic sequences and then predicted prostate shape by sequentially exploring the shape prior.
Subsequently, to mitigate the bias caused by different serialization approaches, they employed a multi-view fusion strategy to merge shape predictions obtained from various perspectives.
Furthermore, they integrated the RNN core into a multi-scale automatic context scheme to continuously refine the details of the shape prediction graph.
Similarly, \cite{seg78} not only utilized CNNs to extract spatial features but also employed RNNs to leverage the temporal information among a series of TRUS images.
Additionally, they utilized cyclic connections within and between different layers of the network to maximize the utilization of temporal information.

\textbf{(c) Attention Enhanced Network.}
In DL methods, an attention block can selectively modify inputs or assign different weights to input variables according to their varying importance.
Attention mechanisms in DL are fundamentally analogous to the selective visual attention mechanisms in humans, with the core objective being to select the most critical information from a vast amount of data for the current task's goal.

In the context of TRUS segmentation, \cite{seg77} firstly developed a network equipped with deep attentional feature (DAF) modules to enhance segmentation by fully exploiting the complementary information encoded in different layers of the CNN.
The DAF module selectively utilized multi-level features integrated across different layers to refine each individual layer, suppressing non-prostatic noise in shallow layers of the CNN, and enriching deeper features with more prostatic detail.
Further, \cite{seg83} extended~\citep{seg77} to 3D TRUS segmentation, by using more powerful backbone architecture.
\cite{seg92} presented a multi-task DL method for boundary detection in intraoperative TRUS, by leveraging both low-level and high-level information.
The channel-wise attention strategy was employed for low-level feature refinement, and the learning-based prior knowledge was utilized for high-level feature modeling.
\cite{seg97} addressed the limitations of transfer learning and fine-tuning methods by introducing a 2.5D deep neural network, which further relied on an attention module that considered feature positioning information to improve segmentation accuracy.
\cite{seg122} developed a coarse-to-fine architecture for TRUS segmentation.
They employed the attention gate module with squeeze-and-excitation U-net to conduct initial segmentation,
and then jointly used a modified polygon-searching model and a storage-based quantum evolution model to refine the segmentation.

Some methods have also explored gland segmentation on micro-ultrasound using Transformer techniques.
MicroSegNet~\citep{seg113}, based on TransUNet~\citep{chen2024transunet}, used multi-scale deep supervision and an annotation-guided loss function to improve segmentation accuracy in difficult regions of the prostate on micro-ultrasound images.
\cite{seg124} employed Swin Transformer~\citep{liu2021swin} as the backbone and further leveraged~\citep{cheng2022masked} to achieve real-time prostate and central gland segmentation on micro-ultrasound images at 9 frames per second.

\subsection{Prostate Image Registration}
\label{subsec:reg}

\begin{figure*}[t]
	\scriptsize
	\[
	\raisebox{-10ex}{\rotatebox{90}{\textbf{Modalities (Registration)}}}
	\begin{cases}
	\text{TRUS-TRUS: }
	\begin{aligned}[t]
	&\textit{\cite{reg6}, \cite{reg9}, \cite{reg13}, \cite{reg22}, \cite{reg28}, \cite{reg40},} \\
	&\textit{\cite{reg44}, \cite{reg53}, \cite{reg56}, \cite{reg64}, \cite{reg65}}
	\end{aligned}\\
	\\
	\text{MRI-TRUS: }
	\begin{aligned}[t]
	&\textit{\cite{reg1}, \cite{reg4}, \cite{reg5}, \cite{reg8}, \cite{reg10},  \cite{reg14}, \cite{reg15},}\\
	&\textit{\cite{reg16}, \cite{reg17}, \cite{reg18}, \cite{reg19}, \cite{reg20}, \cite{reg23},}\\
	&\textit{\cite{reg21}, \cite{reg24}, \cite{reg25}, \cite{reg26}, \cite{reg27}, \cite{reg29},}\\
	&\textit{\cite{reg31}, \cite{reg33}, \cite{reg34}, \cite{reg35}, \cite{reg36}, \cite{reg37}, \cite{reg38},}\\
	&\textit{\cite{reg41}, \cite{reg42}, \cite{reg45}, \cite{reg46}, \cite{reg47}, \cite{reg48},}\\
	&\textit{\cite{reg50}, \cite{reg49}, \cite{reg51}, \cite{reg52}, \cite{reg70},\cite{reg54}, \cite{reg55},} \\
	&\textit{\cite{reg57}, \cite{reg59},\cite{reg61}, \cite{reg63}, \cite{reg67},} \\
	&\textit{\cite{reg68}, \cite{reg69}} \\
	\end{aligned}\\
	\\
	\text{CT-TRUS: }
	\begin{aligned}[t]
	&\textit{\cite{reg3}, \cite{reg7}, \cite{reg32}, \cite{reg39}, \cite{reg58}}\\
	\end{aligned}
	\\
	\end{cases}
	\]
	\caption{Prostate image registration studies categorized as TRUS-TRUS, MRI-TRUS, and CT-TRUS.}
	\label{fig:reg_modality}
\end{figure*}

\begin{figure*}[t]
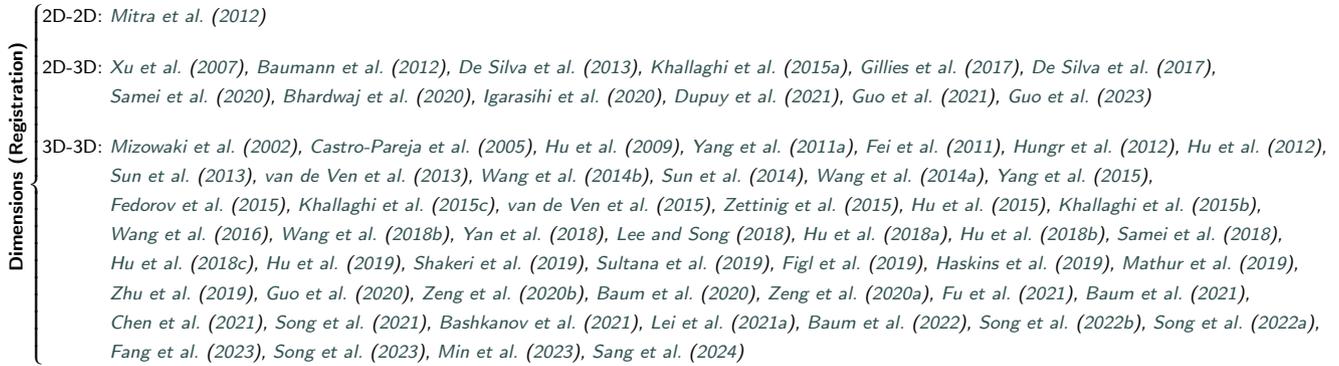

	\scriptsize
	\[
	\raisebox{-10ex}{\rotatebox{90}{\textbf{Dimensions (Registration)}}}
	\begin{cases}
	\text{2D-2D: } \begin{aligned}[t]
	&\textit{\cite{reg8}}
	\end{aligned}\\
	\\
	\text{2D-3D: } \begin{aligned}[t]
	&\textit{\cite{reg4}, \cite{reg11}, \cite{reg13}, \cite{reg22}, \cite{reg28}, \cite{reg30},}\\
	&\textit{\cite{reg45}, \cite{reg46}, \cite{reg50}, \cite{reg53}, \cite{reg56}, \cite{reg64}}
	\end{aligned}\\
	\\
	\text{3D-3D: } \begin{aligned}[t]
	&\textit{\cite{reg1}, \cite{reg3}, \cite{reg5}, \cite{reg6}, \cite{reg7}, \cite{reg9}, \cite{reg10},} \\
	&\textit{\cite{reg14}, \cite{reg15}, \cite{reg16}, \cite{reg17}, \cite{reg18}, \cite{reg19},} \\
	&\textit{\cite{reg21}, \cite{reg20}, \cite{reg23}, \cite{reg24}, \cite{reg25}, \cite{reg26},} \\
	&\textit{\cite{reg27}, \cite{reg29}, \cite{reg31}, \cite{reg32}, \cite{reg33}, \cite{reg34}, \cite{reg35},} \\
	&\textit{\cite{reg36}, \cite{reg37}, \cite{reg38}, \cite{reg39}, \cite{reg40}, \cite{reg41}, \cite{reg42},} \\
	&\textit{\cite{reg44}, \cite{reg47}, \cite{reg48}, \cite{reg49}, \cite{reg51}, \cite{reg52}, \cite{reg70},} \\
	&\textit{\cite{reg54}, \cite{reg55}, \cite{reg57}, \cite{reg58}, \cite{reg59}, \cite{reg61}, \cite{reg63},} \\
	&\textit{\cite{reg65}, \cite{reg67}, \cite{reg68}, \cite{reg69}} \\
	\end{aligned}
	\end{cases}
	\]
	\caption{Prostate image registration studies categorized as 2D-2D, 2D-3D, and 3D-3D.}
	\label{fig:reg_dimension}
\end{figure*}

\subsubsection{Modalities}
Prostate image registration is useful for assisting surgeons during preoperative planning and intraoperative surgery.
It can provide surgeons with more complementary information, especially the multi-modality registration.
Fig.~\ref{fig:reg_modality} lists the categorization of registration approaches according to the image modality.
The majority of studies focused on MRI-TRUS registration, which registered the preoperative MRI scans to the TRUS images, providing better resolution and soft tissue contrast to assist targeted biopsy.
In addition, CT-TRUS registration methods have also been investigated, mainly in the scenarios of treatment planning and prognosis evaluation.
Other studies payed attention to the registration of preoperative and intraoperative TRUS images.

\subsubsection{Dimensions}
We classify registration methods into three categories based on dimensionality: 2D-2D, 2D-3D, and 3D-3D (see Fig.~\ref{fig:reg_dimension}).
Images acquired intraoperatively are predominantly 2D TRUS, whereas those acquired preoperatively are mostly 3D (MRI, CT, 3D TRUS).
Ultimately, the registration methods intended for prostate biopsy would necessitate the registration between intraoperative 2D TRUS and preoperative 3D images.

The process of registering 2D TRUS to 3D images from different modalities presents two main challenges: differing dimensions and distinct modalities.
There was a scarcity of literature that addressed both of these issues simultaneously.
Consequently, most 2D-3D studies concentrated on registration between TRUS images to bridge the dimensional gap,
while 3D-3D studies primarily focused on registration between TRUS and other modalities to overcome the challenge posed by different imaging modalities. 

\begin{figure*}[t]
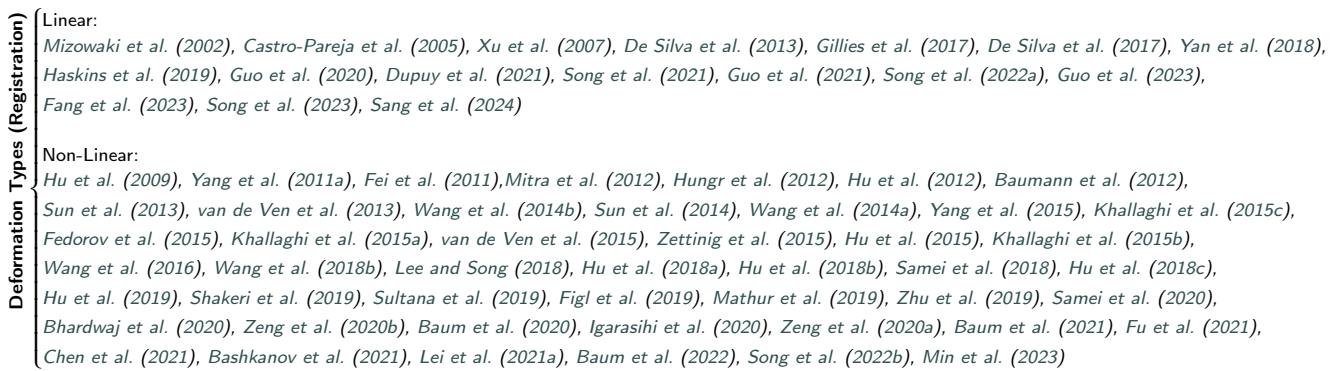

	\scriptsize
	\[
	\raisebox{-14ex}{\rotatebox{90}{\textbf{Deformation Types (Registration)}}}
	\begin{cases}
	\text{Linear:}\\
	\begin{aligned}[t]
	&\textit{\cite{reg1}, \cite{reg3}, \cite{reg4}, \cite{reg13}, \cite{reg28}, \cite{reg30}, \cite{reg31},}\\
	&\textit{\cite{reg41}, \cite{reg47}, \cite{reg53}, \cite{reg55}, \cite{reg56}, \cite{reg63}, \cite{reg64},} \\
	&\textit{\cite{reg65}, \cite{reg67}, \cite{reg69}}\\
	\end{aligned}\\
	\\
	\text{Non-Linear:}\\
	\begin{aligned}[t]
	&\textit{\cite{reg5}, \cite{reg6}, \cite{reg7},\cite{reg8}, \cite{reg9}, \cite{reg10}, \cite{reg11},}\\
	&\textit{\cite{reg14}, \cite{reg15}, \cite{reg16}, \cite{reg17}, \cite{reg18}, \cite{reg19}, \cite{reg20},} \\
	&\textit{\cite{reg21}, \cite{reg22}, \cite{reg23}, \cite{reg24}, \cite{reg25},  \cite{reg26},} \\
	&\textit{\cite{reg27}, \cite{reg29}, \cite{reg32}, \cite{reg33}, \cite{reg34}, \cite{reg35}, \cite{reg36},} \\
	&\textit{\cite{reg37}, \cite{reg38}, \cite{reg39}, \cite{reg40}, \cite{reg42}, \cite{reg44}, \cite{reg45},} \\
	&\textit{\cite{reg46}, \cite{reg48}, \cite{reg49}, \cite{reg50}, \cite{reg51}, \cite{reg70}, \cite{reg52},} \\
	&\textit{\cite{reg54}, \cite{reg57}, \cite{reg58}, \cite{reg59}, \cite{reg61}, \cite{reg68}}\\
	\end{aligned}
	\end{cases}
	\]
	\caption{Prostate image registration studies categorized as linear registration and non-linear registration.}
	\label{fig:reg_deformation}
\end{figure*}

\subsubsection{Deformation Types}
According to different deformation types, we categorize the registration methods into linear and non-linear registration, as shown in Fig.~\ref{fig:reg_deformation}.
Linear transformation, also referred to as isometric transformation, involves only translation and rotation.
Affine transformation includes scaling and shear in addition to translation and rotation, making linear transformation a special case of affine transformation. 
In this survey, both affine and linear transformation are uniformly classified as the category of linear registration.
In contrast, non-linear registration refers to methods that go beyond linear registration, primarily involving deformable transformation.
Unlike linear registration, deformable registration methods employ non-linear dense transformation to align images.
Most of prostate image registration methods focused on non-linear registration to address the issue of large deformation in TRUS images.

\begin{figure*}[t]
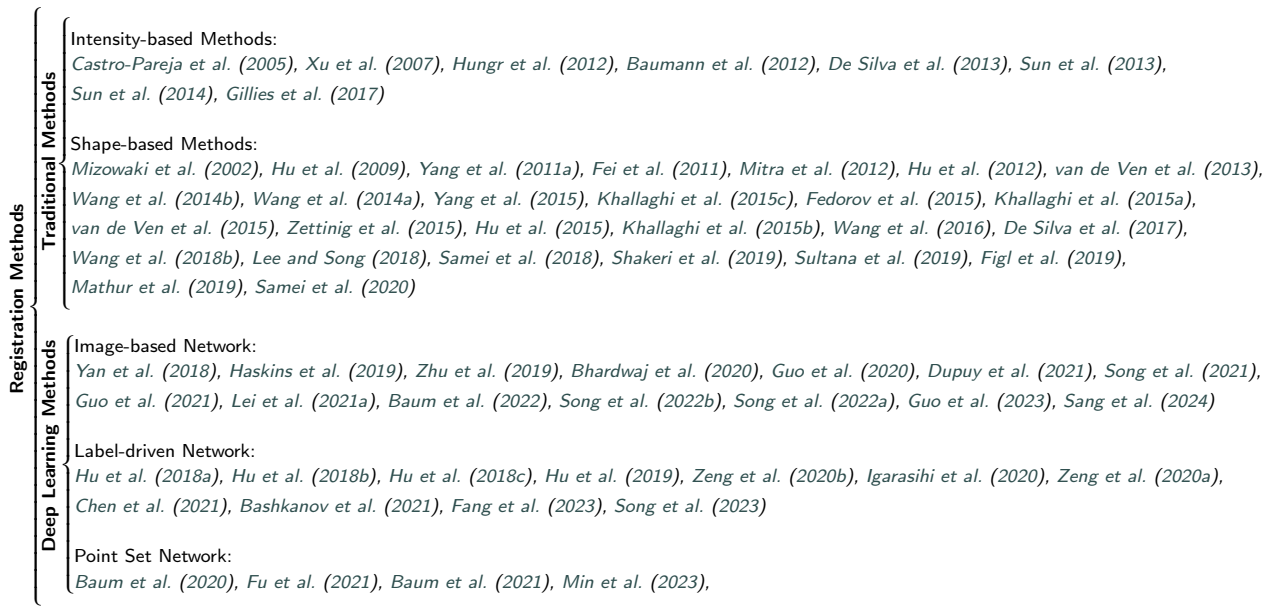

	\scriptsize
	\[
	\raisebox{-10ex}{\rotatebox{90}{\textbf{Registration Methods}}}
	\begin{cases}
	\raisebox{-10ex}{\rotatebox{90}{\textbf{Traditional Methods}}}
	\begin{cases}
	\text{Intensity-based Methods:}\\
	\begin{aligned}[t]
	&\textit{\cite{reg3}, \cite{reg4}, \cite{reg9}, \cite{reg11}, \cite{reg13}, \cite{reg14},} \\
	&\textit{\cite{reg17}, \cite{reg28}} \\
	\end{aligned} 
	\\ \\
	\text{Shape-based Methods:}\\
	\begin{aligned}[t]
	&\textit{\cite{reg1}, \cite{reg5}, \cite{reg6}, \cite{reg7}, \cite{reg8}, \cite{reg10}, \cite{reg15},} \\
	&\textit{\cite{reg16}, \cite{reg18}, \cite{reg19}, \cite{reg20}, \cite{reg21}, \cite{reg22},} \\
	&\textit{\cite{reg23}, \cite{reg24}, \cite{reg25}, \cite{reg26}, \cite{reg27}, \cite{reg30},} \\
	&\textit{\cite{reg29}, \cite{reg32}, \cite{reg35}, \cite{reg38}, \cite{reg39}, \cite{reg40},} \\
	&\textit{\cite{reg42}, \cite{reg45}}
	\end{aligned} 
	\\ 
	\end{cases}
	\\ \\
	\raisebox{-10ex}{\rotatebox{90}{\textbf{Deep Learning Methods}}}
	\begin{cases}
	\text{Image-based Network:}\\
	\begin{aligned}[t]
	&\textit{\cite{reg31}, \cite{reg41}, \cite{reg44}, \cite{reg46}, \cite{reg47}, \cite{reg53}, \cite{reg55},} \\
	&\textit{\cite{reg56}, \cite{reg58}, \cite{reg59}, \cite{reg61}, \cite{reg63}, \cite{reg64}, \cite{reg69}}
	\end{aligned}\\
	\\
	\text{Label-driven Network:}\\
	\begin{aligned}[t]
	&\textit{\cite{reg33}, \cite{reg34}, \cite{reg36}, \cite{reg37}, \cite{reg48}, \cite{reg50}, \cite{reg51},} \\
	&\textit{\cite{reg54}, \cite{reg57}, \cite{reg65}, \cite{reg67}} \\
	\end{aligned}
	\\ \\
	\text{Point Set Network:}\\
	\begin{aligned}[t]
	&\textit{\cite{reg49}, \cite{reg52}, \cite{reg70}, \cite{reg68},}  \\
	\end{aligned}
	\end{cases}
	
	\end{cases}
	\]
	\caption{Traditional registration methods and deep learning (DL) methods. Traditional methods can be categorized as intensity-based methods and shape-based methods. DL methods mainly include image-based networks, label-driven networks, and point set networks.}
	\label{fig:reg_method}
\end{figure*}

\subsubsection{Registration Methods}
We categorize registration methods into two main types: traditional methods and DL methods, as shown in Fig.~\ref{fig:reg_method}.
Traditional methods, while often time-consuming due to the iterative estimation of the deformation field, are known for their stability and reliability across various scenarios.
In contrast, DL methods have gained popularity due to advancements in GPU computing power, enabling them to achieve more efficient registration than traditional methods.

\paragraph{\textbf{Traditional Methods.}}
Traditional registration methods mainly investigated the following two aspects: 
matching criteria and deformation models.
The matching criterion defines the similarity of the image pair, with intensity-based methods relying on the similarity of intensities or intensity-related features of images, and shape-based methods focusing on the similarity of prostate's region/boundary.
The deformation model is also critical, as the registration process involves optimizing the parameters of the deformation fields to warp the moving image to match with the fixed image.
We summarize the traditional registration methods on TRUS images according to different matching criteria.

\textbf{(a) Intensity-based Methods.}
Table~\ref{table:reg_tradition} summarizes some representative traditional registration methods, including intensity-based methods and shape-based methods.
\cite{reg4} employed spatial tracking information to initialize the registration between intraoperative 2D TRUS and preoperative 3D TRUS.
They then optimized the registration by calculating the grayscale similarity to eliminate spatial tracking errors.
Additionally, they manually registered 3D TRUS and 3D MRI preoperatively, achieving a 2D TRUS-3D TRUS-3D MRI registration process.
\cite{reg9} used the correlation coefficient as a similarity metric to register 3D TRUS,
while~\cite{reg11} employed the shift correlation model combined with the inverse consistency constraint, performing linear registration separately from deformable registration.
\cite{reg17} used the modality independent neighborhood descriptor (MIND) as the local similarity feature for MRI-TRUS registration.
They also developed a duality-based convex optimization algorithm to compute MIND similarity.
\cite{reg13} used Powell's method to optimize the normalized cross-correlation (NCC) metric between images, completing the linear  registration of 2D TRUS to 3D TRUS.
Later, \cite{reg28} also employed Powell's method to optimize the NCC metric and realized real-time registration of 2D and 3D TRUS.

\begin{table*}[t]
	\centering
	\caption{Overview of traditional methods for prostate TRUS registration.}
	\fontsize{5}{10}\selectfont
	\begin{tabular}{cccccccc}
		\toprule
		\multirow{2}{*}{\parbox{0.6cm}{\centering Traditional\\Methods}} & 
		\multirow{2}{*}{\centering References} & 
		\multirow{2}{*}{\parbox{0.8cm}{\centering Dimensions\\Modalities}}  & 
		\multirow{2}{*}{\centering Deformation}  & 
		\multirow{2}{*}{\centering Methods} & 
		\multirow{2}{*}{\centering Samples} & 
		\multicolumn{2}{c}{Results} \\
		\cmidrule{7-8} 
		&&&&&& Metrics & Values: final (initial)\\
		\midrule
		
		\multirow{9}{*}{\centering Intensity}
		
		& \cite{reg11} & \parbox{0.8cm}{\centering 2D TRUS\\3D MRI}
		& Non-linear & \parbox{4.5cm}{ 1) Shift correlation and inverse consistency constraint \\ 2) Local rigid and linear elastic deformation} & 40 & TRE & \parbox{2.5cm}{\centering $0.8\pm0.5$ ($13.8\pm7.9$) $\text{mm}$}\\
		\cmidrule{7-8}
			
		& \cite{reg13} & \parbox{0.8cm}{\centering 2D TRUS\\3D TRUS} & Linear & \parbox{4.5cm}{ 1) Optimize the normalized cross-correlation (NCC) metric using Powell's method}
		& 8 & TRE & \parbox{2.5cm}{\centering $1.87\pm0.81$ ($4.75\pm2.62$) $\text{mm}$}\\
		\cmidrule{7-8}
		
		& \cite{reg17} & \parbox{0.8cm}{\centering 3D TRUS\\3D MRI} & Non-linear & \parbox{4.5cm}{ 1) Modality independent neighborhood descriptor \\ 2) Duality-based convex optimization algorithm}
		& 20 & \parbox{0.8cm}{\centering TRE \\ DSC \\ MAD \\ MAXD } & \parbox{2.5cm}{\centering $1.93\pm0.73$ ($3.37\pm1.23$) $\text{mm}$ \\ $85.7\pm4.7$ (n/a) $\%$ \\ $1.84\pm0.52$ (n/a) $\text{mm}$ \\ $6.90\pm2.07$ (n/a) $\text{mm}$}\\
		\cmidrule{7-8}
		
		& \cite{reg28} & \parbox{0.8cm}{\centering 2D TRUS\\3D TRUS} & Linear & \parbox{4.5cm}{ 1) Optimize NCC using Powell's method \\ 2) Golden section search technique} & 14 & TRE 
		&\parbox{2.5cm}{\centering $1.40\pm1.18$ ($3.41\pm3.26$) $\text{mm}$}\\
		\midrule
		
		\multirow{20}{*}{\centering Shape}
	
		
		& \cite{reg8} & \parbox{0.8cm}{\centering 2D TRUS\\2D MRI} 
		& Non-linear & \parbox{4.5cm}{ 1) Point correspondences obtained from a statistical measure of shape-contexts \\ 2) The radial-basis function of thin-plate splines} & 20 & \parbox{0.8cm}{\centering TRE \\ DSC \\ 95HD} & \parbox{2.5cm}{\centering $1.60\pm1.17$ (n/a) $\text{mm}$ \\ $98.0\pm0.4$ (n/a) $\%$ \\ $1.63\pm0.48$ (n/a) $\text{mm}$} \\
		\cmidrule{7-8}
		
		& \cite{reg10} & \parbox{0.8cm}{\centering 3D TRUS\\3D MRI} & Non-linear & \parbox{4.5cm}{ 1) Statistical motion model \\ 2) Expectation maximization algorithm for model-to-image registration}
		& 8 & TRE &\parbox{2.5cm}{\centering $2.40$ ($8.13$) $\text{mm}$}\\
		\cmidrule{7-8}
		
		& \cite{reg24} & \parbox{0.8cm}{\centering 3D TRUS\\3D MRI} & Non-linear & \parbox{4.5cm}{ 1) Coherent point drift algorithm \\ 2) Thin-plate splines} & 13 & TRE &\parbox{2.5cm}{\centering $2.00$ ($2.60$) $\text{mm}$}\\
		\cmidrule{7-8}
		
		& \cite{reg25} & \parbox{0.8cm}{\centering 3D TRUS\\3D MRI} & Non-linear & \parbox{4.5cm}{ 1) Statistical motion model \\ 2) Kernel regression analysis} & 44 & TRE &\parbox{2.5cm}{\centering $2.40$ ($6.19$) $\text{mm}$}\\
		\cmidrule{7-8}
		
		& \cite{reg20} & \parbox{0.8cm}{\centering 3D TRUS\\3D MRI} & Non-linear & \parbox{4.5cm}{ 1) Finite element model with biomechanical constraint \\ 2) Gaussian mixture model}
		& 19 & TRE & \parbox{2.5cm}{\centering $2.72\pm1.15$ ($4.01\pm1.45$) $\text{mm}$}\\
		\cmidrule{7-8}
		
		& \cite{reg26} & \parbox{0.8cm}{\centering 3D TRUS\\3D MRI} & Non-linear & \parbox{4.5cm}{ 1) Statistical shape model \\ 2) Finite element model} & 19 & TRE & \parbox{2.5cm}{\centering $2.35\pm0.81$ ($4.01\pm1.45$) $\text{mm}$}\\
		\cmidrule{7-8}
		
		& \cite{reg27} & \parbox{0.8cm}{\centering 3D TRUS\\3D MRI} & Non-linear & \parbox{4.5cm}{ 1) Personalized statistical deformable model \\ 2) Hybrid point matching} & 18 & TRE & \parbox{2.5cm}{\centering $1.44\pm0.38$ ($6.78\pm1.52$) $\text{mm}$} \\
		\cmidrule{7-8}
		
		& \cite{reg29} & \parbox{0.8cm}{\centering 3D TRUS\\3D MRI} & Non-linear & \parbox{4.5cm}{ 1) Robust projective dictionary learning \\ 2) Personalized statistical deformable model} & 18 & TRE & \parbox{2.5cm}{\centering $1.51\pm0.71$ ($6.78\pm1.52$) $\text{mm}$} \\
		\cmidrule{7-8}
		
		& \cite{reg35} & \parbox{1.0cm}{\centering 2.5D TRUS\\3D MRI}
		& Non-linear & \parbox{4.5cm}{ 1) Linear stress-strain quasi-static FEM \\ 2) 2.5D-3D non-rigid registration} & 11 & TRE & \parbox{2.5cm}{\centering $0.72\pm0.45$ ($4.62\pm3.38$) $\text{mm}$}\\
		
		\bottomrule
	\end{tabular}
	\label{table:reg_tradition}
\end{table*}

\textbf{(b) Shape-based Methods.}
It remains challenging to use intensity-based metrics to evaluate the similarity of multi-modality images.
Therefore, most of MRI-TRUS and CT-TRUS registration approaches sought the shape-based similarities as the solution.
Although the shape-based methods required additional steps to obtain the prostate region or corresponding landmarks, they offered more reliable and robust registration for multi-modality tasks.
\cite{reg8} introduced a non-linear thin-plate spline (TPS) regularized 2D MRI-TRUS registration method.
The boundary point correspondences were estimated based on a statistical measure of shape-contexts.
\cite{reg24} applied the coherent point drift (CPD) algorithm to elastically align surfaces extracted from TRUS and MRI.
They then propagated the surface deformation to the whole gland using the TPS interpolation.
\cite{reg32} proposed a deformable CT-TRUS registration method based on the structural descriptor map (SDM), calculated by solving the Laplace's equation using the segmented prostate surface.
The SDM represented the surface distribution and could be employed for deformable registration.
\cite{reg35} proposed an efficient algorithm to extract arbitrary slices in real-time from deformed 3D images represented by a discrete grid.
They simulated tissue deformation using a patient-specific finite element model (FEM) with boundary conditions for image-guided, robot-assisted laparoscopic radical prostatectomy (RALRP).
Later, \cite{reg45} incorporated this technology into a virtual reality-based RALRP system.

Certain methods~\citep{reg5, reg10} employed both statistical shape modeling (SSM) and biomechanical modeling to capture the motion of the prostate, which has been proved as a robust solution for MRI-TRUS registration.
leveraging the SSM and biomechanical modeling, \cite{reg10} further calculated the normal vector field of the prostate surface in TRUS and used the expectation maximization (EM) algorithm to maximize the log-likelihood function for model-to-image registration.
\cite{reg25} later used kernel regression analysis to express the multivariate subject-specific probability density function for the construction of subject-specific statistical motion model.
\cite{reg20} used a probabilistic framework based on Gaussian mixture model (GMM) to combine FEM with biomechanical prior knowledge for extrapolating and constraining deformation fields during MRI-TRUS registration.
\cite{reg26} employed probabilistic SSM-FEM to map two surfaces and their interiors to a common intermediate SSM instance, aligning MRI to 3D TRUS through this intermediate shape.
\cite{reg27} introduced a hybrid point matching method to establish reliable surface point correspondences.
A series of FEM were conducted using personalized anatomical and biomedical information.
And a personalized statistical deformable model was constructed by the principal component analysis (PCA) of FEMs to estimate the deformation.
To enhance the efficacy of statistical shape modeling in~\citep{reg27}, \cite{reg29} developed a robust projective dictionary learning algorithm to integrate dimensionality reduction and dictionary learning into a unified shape prior modeling framework, and applied it for MRI-TRUS registration.

\paragraph{\textbf{Deep Learning Methods.}}
With the rapid development of GPUs, using DL methods to solve non-linear registration with high computational complexity has become mainstream.
These DL methods for TRUS image registration can be categorized into image-based networks, label-driven networks, and point set networks.
The first two types often take images as input, while the last one uses surface point clouds.
Image-based networks mainly rely on the similarity between moving and fixed images as the loss function to update network parameters.
In contrast, label-driven networks directly use prostate annotations or landmarks to aid registration, either by incorporating them into the loss function or by directly registering them.
Point set networks transform the image registration problem in the form of point cloud matching and train the network using the point-based loss function.
Table~\ref{table:reg_dl} introduces the representative methods in each category.

\begin{table*}[t]
	\centering
	\caption{Overview of deep Learning methods for prostate TRUS registration.}
	\fontsize{5}{10}\selectfont
	\begin{tabular}{cccccccc}
		\toprule
		\multirow{2}{*}{\parbox{0.6cm}{\centering Deep\\Learning\\Methods}} & 
		\multirow{2}{*}{\centering References} & 
		\multirow{2}{*}{\parbox{0.8cm}{\centering Dimensions\\Modalities}}  & 
		\multirow{2}{*}{\centering Deformation}  & 
		\multirow{2}{*}{\centering Methods} & 
		\multirow{2}{*}{\centering Samples} & 
		\multicolumn{2}{c}{Results} \\
		\cmidrule{7-8} 
		&&&&&& Metrics & Values: final (initial)\\
		\midrule
		
		\multirow{10}{*}{\parbox{0.6cm}{\centering Image-based\\Network}}
		
		& \cite{reg31} & \parbox{0.8cm}{\centering 3D TRUS\\3D MRI} & Linear & \parbox{4.5cm}{ 1) Adversarial learning \\ 2) Supervised-learning using ground truth transformation} & 763 & TRE & \parbox{2.5cm}{\centering $3.48$ ($6.11$) $\text{mm}$}\\
		\cmidrule{7-8}
		
		& \cite{reg41} & \parbox{0.8cm}{\centering 3D TRUS\\3D MRI} & Linear & \parbox{4.5cm}{ 1) Use CNNs to learn similarity metrics for MRI–TRUS \\ 2) Composite optimization strategy} & 679 & TRE & \parbox{2.5cm}{\centering $3.86$ ($16$) $\text{mm}$}\\
		\cmidrule{7-8}
		
		& \cite{reg47} & \parbox{0.8cm}{\centering 3D TRUS\\3D MRI} & Linear & \parbox{4.5cm}{ 1) Coarse-to-fine multi-stage framework \\ 2) Generate different training data distributions} & 679 & TRE &\parbox{2.5cm}{\centering $3.57\pm2.03$ ($8$) $\text{mm}$}\\
		\cmidrule{7-8}
		
		& \cite{reg55} & \parbox{0.8cm}{\centering 3D TRUS\\3D MRI} & Linear & \parbox{4.5cm}{ 1) Cross-modal attention mechanism} & 662 & TRE & \parbox{2.5cm}{\centering $3.63\pm1.86$ ($8$) $\text{mm}$}\\
		\cmidrule{7-8}
		
		& \cite{reg56} & \parbox{0.8cm}{\centering 2D TRUS\\3D TRUS} & Linear & \parbox{4.5cm}{ 1) Dual-branch feature extraction module \\ 2) Differentiable 2D slice sampling module} & 619 & TRE & \parbox{2.5cm}{\centering $2.73$ ($5.86$) $\text{mm}$}\\
		\cmidrule{7-8}
		
		
		& \cite{reg64} & \parbox{0.8cm}{\centering 2D TRUS\\3D TRUS} & Linear & \parbox{4.5cm}{ 1) Frame-to-slice correction network \\ 2) Similarity filtering mechanism} & 618 & TRE & \parbox{2.5cm}{\centering $6.95\pm8.03$ (n/a) $\text{mm}$}\\
		\midrule
		
		\multirow{7}{*}{\parbox{0.6cm}{\centering Label-driven\\Network}} 
		
		& \cite{reg33}  & \parbox{0.8cm}{\centering 3D TRUS\\3D MRI} & Non-linear & \parbox{4.5cm}{ 1) Weakly-supervised anatomical-label-driven network \\ 2) Adversarial deformation regularization} & 108 & \parbox{0.8cm}{\centering TRE \\ DSC} & \parbox{2.5cm}{\centering $6.3$ (n/a) $\text{mm}$ \\ $82$ (n/a) $\%$}\\
		\cmidrule{7-8}
		
		& \cite{reg34} & \parbox{0.8cm}{\centering 3D TRUS\\3D MRI} & Non-linear & \parbox{4.5cm}{ 1) Label-driven weakly-supervised learning} & 111 & \parbox{0.8cm}{\centering TRE \\ DSC} & \parbox{2.5cm}{\centering $4.2$ (n/a) $\text{mm}$ \\ $88$ (n/a) $\%$}\\
		\cmidrule{7-8}
		
		& \cite{reg36} & \parbox{0.8cm}{\centering 3D TRUS\\3D MRI} & Non-linear & \parbox{4.5cm}{ 1) Multiscale Dice for weakly-supervised training \\ 2) Memory-efficient network architecture} & 108 & \parbox{0.8cm}{\centering TRE \\ DSC } & \parbox{2.5cm}{\centering $3.6$ (n/a) $\text{mm}$ \\ $87$ (n/a) $\%$} \\
		\cmidrule{7-8}
		
		& \cite{reg37} & \parbox{0.8cm}{\centering 3D TRUS\\3D MRI} & Non-linear & \parbox{4.5cm}{ 1) Conditional image segmentation paradigm} & 115 & \parbox{0.8cm}{\centering TRE \\ DSC} & \parbox{2.5cm}{\centering $2.1$ (n/a) $\text{mm}$ \\ $92$ (n/a) $\%$}\\
		
		\midrule
		\multirow{8}{*}{\parbox{0.6cm}{\centering Point\\Set\\Network}} 
		
		& \cite{reg49} & \parbox{0.8cm}{\centering 3D TRUS\\3D MRI} & Non-linear & \parbox{4.5cm}{ 1) Free point transformer network} & 108 & \parbox{0.8cm}{\centering TRE \\ HD}
		& \parbox{2.5cm}{\centering $4.7\pm1.8$ ($5.1\pm1.7$) $\text{mm}$ \\ $6.3\pm1.6$ ($10.7\pm2.6$) $\text{mm}$}\\
		\cmidrule{7-8}
		
		& \cite{reg70} & \parbox{0.8cm}{\centering 3D TRUS\\3D MRI} & Non-linear & \parbox{4.5cm}{ 1) Data-driven learning without heuristic constraints \\ 2) Variable number of points in training and inference} & 108 & \parbox{0.8cm}{\centering TRE \\ HD } & \parbox{2.5cm}{\centering $4.75\pm1.45$ ($5.1\pm1.7$) $\text{mm}$ \\ $6.18\pm1.36$ ($10.7\pm2.6$) $\text{mm}$}\\
		\cmidrule{7-8}
		
		& \cite{reg52} & \parbox{0.8cm}{\centering 3D TRUS\\3D MRI} & Non-linear & \parbox{4.5cm}{ 1) Biomechanical constraints via finite element analysis \\ 2) Surface Chamber distance loss} & 50 & \parbox{0.8cm}{\centering TRE \\ DSC \\ HD } & \parbox{2.5cm}{\centering $1.57\pm0.77$ (n/a) $\text{mm}$ \\ $94\pm2$ (n/a) $\%$ \\ $2.96\pm1.00$ (n/a) $\text{mm}$}\\
		\cmidrule{7-8}
		
		& \cite{reg68} & \parbox{0.8cm}{\centering 3D TRUS\\3D MRI} & Non-linear & \parbox{4.5cm}{ 1) Biomechanical constraints exerted from soft-tissue-modeling partial differential equations} & 77 & \parbox{0.8cm}{\centering TRE} & \parbox{2.5cm}{\centering $1.90\pm0.52$ (n/a) $\text{mm}$}\\
		
		\bottomrule
	\end{tabular}
	\label{table:reg_dl}
\end{table*}

\textbf{(a) Image-based Network.}
Some methods have focused on the development of novel network architectures.
\cite{reg31} proposed an adversarial MRI-TRUS registration framework, which trained two CNNs concurrently - a generator for image registration and a discriminator to assess the registration quality.
This method only considered linear transformation and required ground truth transformation for the supervised learning.
\cite{reg56} proposed a 2D-3D TRUS registration network.
This network used a dual-branch encoding module to extract 2D and 3D features for the estimation of linear transformation parameters.
And a 2D plane sampling module was designed to enable efficient training and inference.
This approach was subsequently extended to~\citep{reg64},
by newly designing a frame-to-slice correction network to refine the transformation estimation and a similarity filtering mechanism to assess the registration quality.
\cite{reg55} introduced a cross-modal attention module to explicitly utilize spatial correspondence for a linear transformation in 3D MRI-TRUS registration, and further extended this approach to other applications~\citep{reg61}.
\cite{reg69} divided their model into three stages, each of which split and merged fixed and moving images, then fed them into an encoder based on Swin Transformer~\citep{liu2021swin} to obtain an affine transformation for MRI-TRUS.

Other methods have explored various training strategies and loss functions.
\cite{reg41} employed a CNN to learn similarity metrics for the representation of MRI-TRUS, and then used the learned similarity metric to optimize the registration.
\cite{reg47} utilized data generation methods to create diverse training data distributions, allowing the network to perform well for MRI-TRUS registration.
\cite{reg63} introduced the transformed grid distance loss, which addressed the issue of rotation representation and bridged the gap between translation and rotation.
Combining 3D diffusion models~\citep{ho2020denoising} and GANs, \cite{ma2024pmt} translated MRI and US images into an intermediate pseudo-modality, preserving details while generating more similar textures for registration. However, this study did not report the accuracy of the registration.

\textbf{(b) Label-driven Network.}
These networks leveraged prostate annotations or landmarks to evaluate the similarity of multi-modality prostate images.
\cite{reg33} optimized a discriminator network to identify the deformation field predicted by registration network from FE-simulated motion data.
The MRI-TRUS registration network simultaneously aimed to maximize similarity between anatomical labels to drive image alignment and minimize an adversarial generator loss that measured divergence between the predicted and simulated deformation.
\cite{reg34} also investigated using various types of identifiable corresponding anatomical structures to train the network for the prediction of dense voxel correspondence.
These anatomical structures could be prostate region, vessels, point landmarks or other ad hoc structures.
Later, \cite{reg36} extended the work~\citep{reg34},
by designing a new multiscale Dice for the network training and an improved memory-efficient network architecture.
\cite{reg37} applied segmentation concepts to registration, predicting the transformed region-of-interest (ROI) directly without predicting the deformation field.
\cite{reg48} employed a CNN for segmentation and then supervised the training of registration network using a surface loss.
\cite{reg54} also used segmentation-based learning to train the MRI-TRUS registration network for the application of prostate brachytherapy.
\cite{reg67} trained a CNN to extract landmarks and then leveraged these landmarks for affine registration.
\cite{azampour2024multitask} registered MRI to TRUS images via deformable registration, translating MRI into the ultrasound domain while preserving structural content using contrastive unpaired translation.

\textbf{(c) Point Set Network.}
\cite{reg49} described a point set registration algorithm based on a free point transformer network for points extracted from segmented MRI-TRUS surfaces.
Based on the preliminary work in~\citep{reg49},
\cite{reg70} developed a partial point-set registration network, which allowed non-linear matching without known correspondence or heuristic constraints.
\cite{reg52} utilized tetrahedron meshing to generate volumetric prostate point clouds from segmented masks,
and trained a point cloud matching network using deformation fields generated by FE analysis.
\cite{reg68} developed a physically-informed neural network to impose elastic constraints on the estimated displacements, and formulated an end-to-end registration network training algorithm by minimizing surface distances as estimated boundary conditions in partial differential equations.

\subsection{Prostate Cancer Classification and Detection}
\label{subsec:cla}

\begin{figure*}[t]
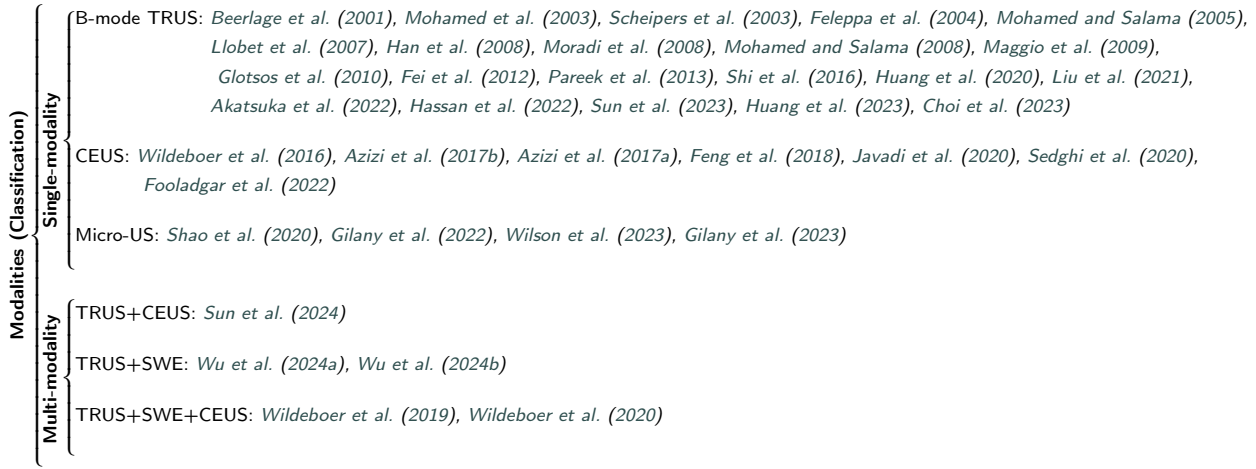

	\scriptsize
	\[
	\raisebox{-12ex}{\rotatebox{90}{\textbf{Modalities (Classification)}}}
	\begin{cases}
	\raisebox{-10ex}{\rotatebox{90}{\textbf{Single-modality}}}
	\begin{cases}
	\text{B-mode TRUS: }
	\begin{aligned}[t]
	&\textit{\cite{beerlage2001correlation}, \cite{mohamed2003prostate}, \cite{scheipers2003ultrasonic}, \cite{feleppa2004ultrasonic}, \cite{mohamed2005computer},}\\
	&\textit{\cite{llobet2007computer}, \cite{han2008computer}, \cite{moradi2008augmenting}, \cite{mohamed2008prostate}, \cite{maggio2009predictive},}\\
	&\textit{ \cite{glotsos2010multi}, \cite{fei2012molecular}, \cite{pareek2013prostate}, \cite{shi2016stacked}, \cite{huang2020texture}, \cite{liu2021deep},}\\
	&\textit{\cite{akatsuka2022data}, \cite{hassan2022prostate}, \cite{sun2023three}, \cite{huang2023transfer}, \cite{choi2023mic}}
	\end{aligned}
	\\ \\
	\text{CEUS: }
	\begin{aligned}[t]
	&\textit{\cite{wildeboer2016multiparametric}, \cite{azizi2017transfer}, \cite{azizi2017detection}, \cite{feng2018deep}, \cite{javadi2020multiple}, \cite{sedghi2020improving},}\\
	&\textit{ \cite{fooladgar2022uncertainty}}
	\end{aligned}                 
	\\ \\  
	\text{Micro-US: } 
	\begin{aligned}[t]
	&\textit{\cite{shao2020improving}, \cite{gilany2022towards}, \cite{wilson2023self}, \cite{gilany2023trusformer}}
	\end{aligned}
	\\ \\  
	\end{cases}
	\\ \\
	\raisebox{-10ex}{\rotatebox{90}{\textbf{\quad Multi-modality}}}
	\begin{cases}
	\text{TRUS+CEUS: }
	\begin{aligned}[t]
	&\textit{\cite{sun2024machine}}
	\end{aligned}
	\\ \\
	\text{TRUS+SWE: }
	\begin{aligned}[t]
	&\textit{\cite{wu2024multi}, \cite{wu2024miccai}}
	\end{aligned}
	\\ \\
	\text{TRUS+SWE+CEUS: }
	\begin{aligned}[t]
	&\textit{\cite{wildeboer2019machine}, \cite{wildeboer2020automated}}
	\end{aligned}
	\\ \\
	\end{cases}
	\\
	\end{cases}
	\]
	\caption{Prostate cancer classification studies categorized as using single-modality and multi-modality.}
	\label{fig:cla_modality}
\end{figure*}

\begin{table*}[t]
	\centering
	\caption{Overview of TRUS-based prostate cancer classification and detection methods.}
	\label{table:cls}
	\resizebox{\textwidth}{!}{
		\begin{tabular}{@{}lllccccc@{}}
			\toprule
			\multirow{2}{*}{Types} & \multirow{2}{*}{References} & \multirow{2}{*}{Methods} & \multirow{2}{*}{Samples} & \multicolumn{4}{c}{Results} \\ 
			\cmidrule{5-8} 
			& & & & Acc (\%) & Sen (\%) & Spe (\%) & AUC (\%)\\
			\midrule
			\multirow{7}{*}{\makecell[l]{Traditional \\ Methods}}
			& \cite{mohamed2005computer} & Texture features, support vector machine (SVM) & 33 & 87.75 & 83.33 & 90.00 & -    \\
			& \cite{han2008computer} & Texture features, clinical features, SVM & 51 & 96.40 & 92.00 & 95.90 & -    \\
			& \cite{moradi2008augmenting} & Texture features, radio frequency (RF) features, SVM & 35 & -     & -     & -     & 95.00 \\
			& \cite{maggio2009predictive} & Multi-feature kernel classification, predictive deconvolution & 37 & 89.00 & 79.00 & 89.00 & 93.00 \\
			& \cite{glotsos2010multi} & Texture features, multi-classifier system, hierarchical decision tree & 165 & 88.70 & 86.70 & 89.80 & -     \\
			& \cite{wildeboer2020automated} & Contrast enhanced features, radiomics features, random forest & 50 & -     & -     & -     & 75.00 \\
			& \cite{sun2024machine} & Clinical risk factors, radiomics, multivariable logistic regression & 166 & 82.00 & 80.00 & 83.00 & 89.00 \\
			\midrule
			\multirow{11}{*}{\makecell[l]{Deep \\ Learning \\ Methods}}
			& \cite{shi2016stacked} & Stacked deep polynomial network, representation learning & 70 & 86.11 & 76.40 & 91.45 & -     \\
			& \cite{azizi2017detection} & Deep belief network,  tissue mimicking, Gaussian mixture model & 197 & 68.00 & 63.00 & 67.00 & 72.00 \\
			& \cite{feng2018deep} & 3D convolutional neural network (CNN), spatial-temporal features  & 27841 & 90.18 & 82.83 & 91.45 & -     \\
			& \cite{shao2020improving} & Generative adversarial network, modified U-net & 6607 & - & 95.10 & 87.70 & 93.40 \\
			& \cite{javadi2020multiple} & Multiple instance learning, conditional variational auto encoders & 339 & 66.00 & 77.00 & 55.00 & 68.00 \\
			& \cite{gilany2022towards} & Co-teaching paradigm, evidential deep learning & 472 & -     & 67.38 & 88.20 & 87.76 \\      
			& \cite{gilany2023trusformer} & Transformer, CNN,  self-supervised learning (SSL) & 1398 & -     & 88.00 & 51.00 & 80.00 \\
			& \cite{sun2023three} & 3D CNN, prostate segmentation & 832 & -     & 81.00 & 78.00 & 85.00 \\
			& \cite{wilson2023self} & SSL, pretrain and fine-tune & 1028 & -     & -     & -     & 90.99 \\
			& \cite{wu2024multi} & 3D CNN, adaptive spatial fusion
			module, orthogonal regularization  & 512 & 79.00 & -     & -     & 84.00 \\
			& \cite{wu2024miccai} & 3D CNN, hybrid attention module, prototype correction module & 512 & 81.00 & 86.00 & 73.00 & 86.00 \\
			\bottomrule
		\end{tabular}
	}
\end{table*}

\subsubsection{Modalities}
As depicted in Fig.~\ref{fig:cla_modality}, the predominant modalities for TRUS-based PCa classification and detection methods are B-mode TRUS, as well as contrast-enhanced ultrasound (CEUS) and micro-ultrasound (Micro-US).
With the advent of multi-modality imaging techniques, such as shear wave elastography (SWE), there has been a growing interest in multi-modality TRUS-based PCa diagnosis in recent years.

\subsubsection{Classification and Detection Methods}
We categorize the TRUS-based PCa diagnosis methods into two classes:
traditional methods and deep learning methods.
Some representative studies are summarized in Table~\ref{table:cls}.

\paragraph{\textbf{Traditional Methods.}}
These methods have payed attention to analyzing the hand-crafted features extracted from images to conduct diagnosis.
\cite{mohamed2005computer} extracted texture features from TRUS, including the gray level dependence matrix and gray level difference vector.
These features were then ranked and selected using a mutual information based selection algorithm to ensure that only relevant features were used for classification using SVMs.
\cite{han2008computer} presented a method that combined multi-resolution auto-correlation texture features with clinical features, such as tumor location and shape, using a SVM for classification.
\cite{moradi2008augmenting} searched for the best-performing feature subsets from texture features, radio frequency (RF) time series features, and Lizzi-Feleppa (LF) features, and combined these optimal subsets to construct a SVM classifier.
\cite{maggio2009predictive} designed a multi-feature kernel classification model based on generalized discriminant analysis.
They further employed the predictive deconvolution to reduce system-dependent effects.
\cite{glotsos2010multi} developed a multi-classifier system for discriminating between normal, infectious, and cancerous prostate based on the texture analysis of TRUS.
A two-level hierarchical decision tree was designed to first distinguish between normal and abnormal cases, and then to further classify the abnormal cases into infectious and cancerous categories.
Classification at each level was conducted using three classifiers - cubic least square mapping probabilistic neural network, quadratic Bayesian, and SVM.
\cite{wildeboer2020automated} constructed a classification model that utilized TRUS, SWE, and CEUS.
Initially, features concerning contrast perfusion and dispersion were extracted from CEUS videos.
Subsequently, radiomics features were collected from all imaging modalities, and a random forest model was employed for classification.
\cite{sun2024machine} collected clinical risk factors, including age, PSA density, serum total PSA, rise time, as well as radiomics features from target lesions identified in B-mode and CEUS.
These features were then used to develop a risk factors-radiomics combined model through multivariable logistic regression analysis.

\begin{figure*}[t]
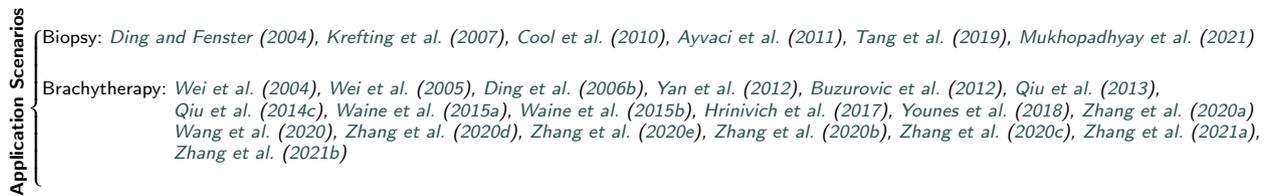

	\scriptsize
	\[
	\begin{aligned}
	\raisebox{-10ex}{\rotatebox{90}{\textbf{Application Scenarios}}}
	\begin{cases}
	\text{Biopsy: } 
	\begin{aligned}[t]     
	&\textit{\cite{ding2004projection}, \cite{krefting2007segmentation}, \cite{cool2010temporal}, \cite{ayvaci2011biopsy}, \cite{tang2019real}, \cite{mukhopadhyay2021deep}}\\
	\end{aligned}
	\\ \\
	\text{Brachytherapy: } 
	\begin{aligned}[t]
	&\textit{\cite{wei2004oblique}, \cite{wei2005oblique}, \cite{ding2006needle}, \cite{yan2012automatic}, \cite{buzurovic2012needle}, \cite{qiu2013needle},}\\
	&\textit{\cite{qiu2014phase}, \cite{waine20153d}, \cite{waine2015three}, \cite{hrinivich2017simultaneous}, \cite{younes2018automatic}, \cite{zhang2020weakly}}\\
	&\textit{\cite{wang2020deep}, \cite{zhang2020multi1}, \cite{zhang2020automatic}, \cite{zhang2020multi3}, \cite{zhang2020multi2}, \cite{zhang2021ultrasound},}\\
	&\textit{\cite{zhang2021multi}}
	\end{aligned} 
	\\ \\
	\end{cases}
	\end{aligned}
	\]
	\caption{Needle detection approaches categorized as applications in biopsy and brachytherapy.}
	\label{fig:det_application}
\end{figure*}

\paragraph{\textbf{Deep Learning Methods.}}
The DL methods for TRUS-based PCa diagnosis were primarily aimed at achieving a more powerful feature representation that enhanced the classification performance.
\cite{shi2016stacked} employed a deep polynomial network (DPN) to improve the representation performance of the initially extracted texture features for small TRUS dataset, and further proposed a stacked DPN algorithm to enhance the original DPN.
\cite{azizi2017detection} employed deep belief networks to automatically learn high-level features from CEUS data for PCa detection and grading.
Additionally, they utilized GMMs to model the distribution of these features across different Gleason grades.
\cite{feng2018deep} proposed a 3D CNN to extract the spatial-temporal features uniformly from the CEUS videos.
The video frames were split into small image tensors, and each tensor was treated as a sample.
\cite{shao2020improving} utilized a generative adversarial network (GAN)-based three-player minimax game framework to address data source heterogeneity and enhance classification performance for PCa. 
A modified U-net was proposed to encode multi-scale PCa-related information, thereby facilitating more effective data analysis and classification.
\cite{javadi2020multiple} employed multiple instance learning (MIL) networks to learn from TRUS regions corresponding to biopsy core pathology, which represented a statistical distribution of PCa. 
Independent conditional variational auto encoders were trained to extract label-invariant features from RF data, facilitating the generation of synthetic data that enhanced MIL network training.
\cite{gilany2022towards} proposed a deep learning model for the detection of PCa using micro-US.
They employed a co-teaching paradigm to handle label noise and incorporated evidential DL to estimate predictive uncertainty.
This combined approach addressed challenges such as weak labels and out-of-distribution data.
\cite{gilany2023trusformer} proposed a multi-scale and context-aware DL method for PCa detection from micro-US.
Self-supervised learning (SSL) was used to train a CNN for extracting features from ROIs, while a Transformer model with attention mechanisms was employed for feature aggregation.
\cite{sun2023three} proposed a mask-guided hierarchical 3D CNN for identifying clinically significant prostate cancer (csPCa) from TRUS videos.
\cite{wilson2023self} employed a SSL approach to leverage abundant unlabeled micro-US data to train the PCa detection model.
By pretraining models with SSL and fine-tuning them on limited labeled data, the approach outperformed traditional supervised learning methods.
\cite{wu2024multi} proposed a framework for the classification of csPCa from multi-modality TRUS videos.
The framework utilized two encoders to extract features from B-mode and SWE.
An adaptive spatial fusion module was introduced to aggregate two modalities' features.
An orthogonal regularized loss was further used to mitigate feature redundancy.
\cite{wu2024miccai} further designed a hybrid attention module to aggregate features from B-mode and SWE, then employed a prototype correction module to refine these features for better classification of csPCa from multi-modality TRUS videos.
\cite{choi2023mic} learned ultrasound biomarkers by leveraging non-registered multimodal image correlations (ultrasound, MRI, histopathology), feeding enriched features into a 3D-UNet for PCa detection.

 

\subsection{Needle Detection}
\label{subsec:det}
\subsubsection{Biopsy and Brachytherapy}
As shown in Fig.~\ref{fig:det_application}, most detection/segmentation approaches for interventional needles from TRUS have been focused on the clinical scenarios of biopsy and brachytherapy procedures.
In the biopsy, needles are inserted to obtain targeted tissue samples, which are analyzed to identify and diagnose PCa.
In the brachytherapy, needles are used to implant radioactive seeds directly into cancerous tissues, maximizing treatment efficacy while minimizing damage to healthy areas.

\begin{table*}[t]
	\centering
	\caption{Overview of methods on needle detection from TRUS.}
	\resizebox{\textwidth}{!}{
		\begin{tabular}{@{}llccccc@{}}
			\toprule
			\multirow{2}{*}{References} & \multirow{2}{*}{Methods} & \multirow{2}{*}{Samples} & \multicolumn{3}{c}{Results} \\ 
			\cmidrule{4-6} 
			& & &  Shaft Error ($\text{mm}$) & Shaft Error ($^{\circ}$) & Tip Error ($\text{mm}$) \\
			\midrule
			\cite{wei2005oblique} & Grey-level change detection, thresholding, linear regression & Phantom & -             & 0.52$\pm$0.16 & 0.82$\pm$0.11 \\
			
			\cite{cool2010temporal} & Hough transform, temporal information & 108 & - & 2.30$\pm$2.00 & 2.10$\pm$4.00 \\
			
			\cite{qiu2013needle} & 3D Hough transform & 35 & -             & 0.80$\pm$0.21 & 0.90$\pm$0.32 \\
			
			\cite{qiu2014phase} & 3D phase grouping technique, 3D Hough transform & 50 & - & 0.95$\pm$0.15 & 1.15$\pm$0.15 \\
			
			\cite{waine2015three} & Random sample consensus (RANSAC) & Phantom & 0.37$\pm$0.15 & - & - \\
			
			\cite{zhang2020weakly} & Bidirectional convolutional sparse coding, RANSAC & 10 & 0.11$\pm$0.09 & - & 0.32$\pm$0.35 \\
			 
			\cite{zhang2020automatic} & Large margin mask R-CNN & 23 & 0.09$\pm$0.04 & - & 0.32$\pm$0.30 \\
			
			\cite{zhang2020multi2} &  Unsupervised order-graph regularized sparse dictionary learning & 91 & 0.19$\pm$0.13 & - & 1.01$\pm$1.74 \\ 
			
			
			\bottomrule
		\end{tabular}
	}
	\label{table:det}
\end{table*}

\subsubsection{Needle Detection Methods}
We categorize these methods into traditional methods and deep learning methods.
Some representative studies are given in Table~\ref{table:det}.

\paragraph{\textbf{Traditional Methods.}}
\cite{wei2005oblique} used grey-level change detection by comparing 3D TRUS captured before and after needle insertion.
The needle was segmented from the difference map using a series of steps including thresholding, removal of spurious signals, and linear regression.
\cite{waine2015three} employed random sample consensus (RANSAC) to reconstruct the 3D needle shape during TRUS-guided prostate brachytherapy.
The Hough transform, traditionally employed in image processing for the identification of geometrical shapes like lines and circles, has been adapted for the segmentation of interventional needles in TRUS.
\cite{cool2010temporal} introduced a temporal-based algorithm that harnessed the Hough transform to detect and track needle edges across multiple 2D TRUS frames.
\cite{qiu2013needle} further employed the 3D Hough transform, leveraging edge detection techniques in conjunction with assembly into coherent structures.
This method facilitated the estimation of needle deflection angles with enhanced precision.
\cite{qiu2014phase} grouped voxel gradient vectors into line support regions (LSRs).
Each LSR was processed using least-squares fitting and a 3D randomized Hough transform to estimate the needle axis.



\paragraph{\textbf{Deep Learning Methods.}}
In recent years, as DL has rapidly advanced, it has also gained increasing prominence in the task of needle detection.
\cite{wang2020deep} developed a modified U-net for segmenting pixels belonging to brachytherapy needles from TRUS, supplemented by an additional VGG-16-based network that predicted the locations of the needle tips.
\cite{zhang2020weakly} adopted a weakly supervised framework that referenced corresponding CT images due to their higher signal-to-noise ratio.
The approach was centered around the bidirectional convolutional sparse coding (BiCSC) model, which concurrently learned dictionaries for both US and CT images to enhance needle visualization and reduce noise.
The whole process included image preprocessing, needle reconstruction using the BiCSC model, and needle modeling through iterative RANSAC algorithms.
\cite{zhang2020automatic} utilized a large margin mask R-CNN model to localize needle shafts in TRUS, and further employed a needle-based density-based spatial clustering of applications with noise (DBSCAN) algorithm to refine shaft localization and detect needle tips.
\cite{zhang2020multi2} integrated sparse dictionary learning with order-graph regularization to enhance spatial continuity in TRUS.
The key idea lied in a label-free approach that leveraged auxiliary images without needles to learn intrinsic features for reconstructing target images.

\section{Discussion and Future Research Directions}
\label{sec:discussion}
In this section, we discuss the current challenges and suggest possible research directions for each of the main tasks in TRUS image processing.

\subsection{Prostate Gland Segmentation}
\label{subsec:dis_seg}
The evolution of TRUS segmentation techniques is pivotal for enhancing the efficiency of PCa diagnosis and treatment.
The intrinsic challenge in TRUS segmentation arises from its imaging characteristics and the physiological properties of the prostate, which collectively make the delineation of prostate boundaries in TRUS ambiguously.
Section~\ref{subsec:seg} provides a narrative review of the research progress in the field of TRUS segmentation, analyzing it from the perspectives of imaging modalities, dimensions, and segmentation methods.
Diverse imaging modalities such as B-mode, elastography, and micro-US offer various structural representations, aiding in the identification of the prostate's morphology and boundaries.
Moreover, 2D, 2.5D, and 3D segmentation methods each have their advantages, ranging from efficient segmentation of single slices to comprehensive analysis of the entire volume, offering a variety of technical options for accurate prostate segmentation.
Traditional segmentation methods and DL-based approaches each have their characteristics, with DL showing great potential in improving segmentation accuracy.

According to Table~\ref{table:seg_trad}, for traditional 2D TRUS gland segmentation methods, prostate segmentation based on principal curves achieves excellent results (DSC $\textgreater$ 95\%), though it is semi-automatic.
Among fully automatic 2D segmentation methods, active shape model-based approaches also yield satisfactory results.
In 3D segmentation, methods leveraging symmetry and convex optimization demonstrate high accuracy (DSC $\textgreater$ 93\%). 
For deep learning methods according to Table~\ref{table:seg_dl}, 2D segmentation using multi-scale strategies achieves high DSC values, while for 3D segmentation, current deep learning methods generally underperform compared to traditional methods.
However, some approaches, such as~\citep{seg109} and~\citep{seg83}, achieve promising results.

Despite significant advancements in TRUS segmentation techniques, multiple challenges remain.
The ambiguity and noise in TRUS images pose a particular difficulty for the automatic segmentation of prostate boundaries.
Moreover, while 3D segmentation methods offer more comprehensive structural information of the prostate, their high computational costs and demands on computational resources limit their application in clinical practice.
The scarcity and inconsistency of data quality also challenge the generalizability of models, especially in surgical navigation and treatment planning, where high-precision segmentation is required.
The demand for real-time performance further complicates the design of segmentation algorithms, particularly in surgical environments where there are stringent requirements for the response speed of the algorithms.

Future research should focus on enhancing the automation, accuracy, and real-time performance of TRUS segmentation techniques.
Automated and intelligent algorithms will reduce user interaction, enhancing the efficiency and repeatability of segmentation.
Continuous optimization of DL models, including improvements in network architecture and training strategies, will better adapt to the specific requirements of prostate segmentation.
Few-shot learning, semi-supervised and weakly-supervised learning methods will address the scarcity of high-quality annotated data, improving model performance with limited data.
Moreover, cross-modality and multi-task learning methods will utilize the correlation between different imaging modalities and related clinical tasks to enhance the overall effectiveness of segmentation. 
Finally, improving the interpretability and generalizability of models, as well as optimizing models to meet the demands of real-time processing, will be key to advancing the development of prostate segmentation technology.

\subsection{Prostate Image Registration}
In the domain of PCa diagnosis and therapy, the registration of TRUS images constitutes a critical technological component.
By aligning preoperative imaging (could be 3D MRI, CT, TRUS) with real-time 2D TRUS obtained during surgery, it facilitates the precise targeting and treatment by clinicians.
We take MRI-TRUS registration for instance.
The prerequisite for this process is the acquisition of MRI, typically T2-weighted multiparametric MRI (mpMRI).
Diagnosis is performed on the MRI to identify biopsy locations. 
During intraoperative MRI-TRUS fusion, there are currently two main approaches:
(1) Cognitive fusion: this method involves the ultrasound operator observing both TRUS and MRI simultaneously to locate the suspicious regions identified on the MRI within the TRUS images. 
While this approach is operator-friendly, it requires a deep understanding of both modalities and the ability to find corresponding locations between them, necessitating an experienced operator.
(2) Image registration-based fusion: this technique uses algorithms to perform rigid or deformable registration of the prostate gland between MRI and TRUS, often involving segmentation of the gland in both modalities.
The registered spatial correspondence is then used to transfer the biopsy target location from the MRI to the TRUS.
This method significantly reduces the reliance on operator experience in image interpretation but requires additional computational resources for running registration and segmentation algorithms, and it must account for potential algorithmic errors.
Conventional image registration approaches mainly include intensity-based methods and shape-based methods, where the shape-based statistical model combined with biomechanical modeling methods provide a robust set of solutions.
Advancing with the evolution of DL, registration methods driven by point clouds or annotated labels have exhibited enhanced automation and precision by discerning intricate mappings between sets of images.

Tables~\ref{table:reg_tradition} and~\ref{table:reg_dl} show that, from a performance perspective, traditional methods generally outperform deep learning methods in terms of registration accuracy (with lower TRE).
It is possible that some iterative strategies used in traditional methods could be integrated into current deep learning approaches to achieve better results, balancing accuracy and real-time performance.
Interestingly, the best-performing methods in both traditional and deep learning categories (\citep{reg35}, \citep{reg68}) utilize FEM, which shows strong research potential by integrating biomechanical modeling to produce more realistic deformations. 
Additionally, intensity-based traditional methods, such as~\citep{reg11}, also achieve favorable results, and their strategies could be incorporated into future deep learning methods.

Regarding registration modalities and types, traditional methods cover a broader range.
In deep learning approaches, linear registration focuses primarily on 3D TRUS-MRI registration and 2D TRUS to 3D TRUS registration.
For non-linear registration, most current deep learning methods address 3D multi-modality registration, often requiring additional segmentation to assist in training, including label-driven and point-set methods.
In the future, deep learning research could expand to broader areas, such as deformable 2D to 3D TRUS registration, rigid 2D TRUS to 3D MRI registration, and deformable 2D TRUS to 3D MRI registration.
Although~\cite{ma2024pmt} employs an unsupervised 3D generation method to generate fake modality for further registration, it does not provide additional details on the registration approach or results.
However, this direction is quite innovative.

According to Tables~\ref{table:reg_tradition} and~\ref{table:reg_dl}, only traditional methods have explored rigid registration from 2D TRUS to 3D MRI, but these methods require iterative processes.
In contrast, current deep learning approaches are limited to rigid registration from 3D to 2D TRUS or deformable registration between 3D MRI and 3D TRUS, with the additional requirement of segmentation.
Thus, using deep learning for registration necessitates acquiring 3D TRUS images from patients or generating 3D TRUS from 3D MRI through generative techniques.
This process involves segmenting and registering the gland in both 3D images, then performing rigid registration of 3D to 2D TRUS intraoperatively, and finally mapping the 2D TRUS slices onto the 3D MRI using deformable registration.
This workflow is currently complex, requiring additional computational resources and software, and is not very convenient for clinical use.

Most of the articles we surveyed involve multi-modality registration, predominantly focusing on 3D-3D registration.
Consequently, shape-based methods or label-driven networks are often employed in these tasks.
These methods are generally unaffected by different modalities and have demonstrated promising results in preoperative applications. 
However, shape-based methods, which register according to the prostate contours, tend to produce sparse deformation fields, leading to sub-optimal deformation effects in internal areas.
Additionally, both shape-based and label-driven methods require segmentation and annotation, which not only suffer from the segmentation challenges discussed in Section~\ref{subsec:dis_seg} but also increase time complexity, rendering them unsuitable for intraoperative scenarios.
Intraoperative registration involves the challenging leap from 3D to 2D dimensions, making it difficult to ensure both accuracy and real-time performance simultaneously, which is why an end-to-end method has yet to be developed.
While image-based methods can quickly generate dense deformation fields, their accuracy needs improvement.
Finding an effective similarity metric to measure different modalities may be key to solving this problem.

Looking ahead to the future development of TRUS image registration, a suite of innovative research directions is anticipated to propel the advancement of this field.
At the forefront, researchers are committed to crafting novel algorithms that integrate detailed internal structural information of the prostate, alongside the design of loss functions capable of surmounting the disparities between various imaging modalities.
An ideal solution would be end-to-end registration, though this is challenging to implement.
Alternatively, generating TRUS from MRI or using an intermediate modality could avoid the need for 3D TRUS acquisition.
However, this requires precise image conversion and ensures that tissue information, especially related to PCa, is preserved.
Additionally, the integration of SWE into the registration process to leverage the biomechanical properties of tissues for improved registration accuracy is an important avenue for future exploration.
Collectively, these research directions aim to foster progress in TRUS image registration, equipping the clinical realm with more accurate diagnostic and therapeutic tools.

\subsection{Prostate Cancer Classification and Detection}
In the realm of PCa classification and detection, B-mode TRUS has emerged as a predominant imaging modality, complemented by CEUS, Micro-US, and SWE to enhance the accuracy of diagnosis.
Traditional methods are primarily relied on texture features, or combined with RF features for analysis through classifiers like SVM.
Moreover, radiomic analysis offers a comprehensive feature assessment, extracting a vast array of quantitative features from images.
The application of DL has revolutionized PCa classification, with various network structures and training strategies significantly improving classification performance.

According to Table~\ref{fig:cla_modality}, the majority of current methods rely on single-modality input, with only five using multi-modality approaches.
Notably, none of the single-modality methods independently consider elastography.
From a clinical perspective, elastography and contrast-enhanced imaging provide important indicators for diagnosing PCa, so there is significant room for research in combining multiple modalities in computer-aided PCa diagnosis.
Table~\ref{table:cls} highlights that, in terms of AUC, deep learning methods have not shown a substantial improvement over traditional methods, despite using larger sample sizes.
Among traditional methods, SVM approaches based on texture features perform exceptionally well.
Combining the texture features from traditional methods with deep learning approaches could be a promising direction for future research.

Several challenges persist in this field.
Data annotation remains a difficult and time-consuming task due to the intricate nature of prostate lesions in TRUS.
Additionally, the heterogeneity of data sources and the limited availability of high-quality labeled data complicate the development of robust models.
These issues underscore the need for innovative solutions to enhance data quality and model performance.

Future research should explore the integration of multi-modality data and advanced learning strategies to overcome these challenges.
By combining various imaging modalities and employing techniques like weakly-supervised and semi-supervised learning, researchers can better utilize the vast amounts of available but unlabeled data.
Such approach can help mitigate the difficulties in data annotation and improve the generalization capabilities of their models. 
Furthermore, while most current studies focus on classification tasks, there is a crucial need to develop methods for lesion-scale detection.
Advancing research in this direction will enable more precise localization and characterization of cancerous regions, enhancing the overall efficacy of PCa diagnostics in clinical practice.

\subsection{Needle Detection}
Recent advancements in interventional needle detection have leveraged both traditional and DL methods.
Traditional techniques, such as the Hough transform and random sample consensus (RANSAC), have been used to detect and track needle edges in TRUS images.
These methods rely on identifying geometrical shapes and managing noise and artifacts effectively.
Meanwhile, DL approaches have revolutionized needle detection by automatically learning complex patterns from large datasets,  improving the accuracy and efficiency of needle detection.

Despite these advancements, several challenges remain in this field.
The poor imaging quality of TRUS often complicates the detection process.
Furthermore, there is often a discrepancy in annotating the gold standard, leading to errors in needle segmentation.
These issues highlight the need for more robust techniques to enhance image quality and reduce annotation errors.

Future research should consider the inherent errors in the gold standard annotations, weakly-supervised and semi-supervised learning methods should be explored to improve model training.
Further advancements could include developing innovative algorithms that can better handle the inherent noise and artifacts in TRUS images, paving the way for more precise and reliable prostate needle segmentation methods.
The integration of these techniques into clinical workflows will further facilitate the delivery of personalized and targeted therapies, contributing to the overall improvement in patient care and outcomes.

\section{Summary}
\label{sec:conclusion}
This paper conducts a comprehensive review of the advancements in transrectal ultrasound image processing.
The paper meticulously reviews a plethora of studies spanning two decades, delineating the paradigm shift from conventional methods to deep learning approaches.
\textit{For the segmentation},
the paper summarizes the traditional shape/contour-based, region-based, machine learning methods, as well as the DL models.
For 3D TRUS segmentation, the exigency for high accuracy in preoperative planning, image registration, and volume measurement is paramount, necessitating solutions to low image resolution and large shape variations.
Conversely, 2D TRUS segmentation confronts the challenge of achieving real-time speed, especially in intraoperative navigation scenarios.
\textit{For the registration},
the paper deliberates on the efficacy of traditional intensity-based and shape-based methods, as well as the burgeoning DL approaches, including image-based, label-driven, and point set methods.
It notes the predominance of multi-modality 3D-3D registration tasks and the challenges associated with real-time property and accuracy in intraoperative settings.
\textit{For the PCa diagnosis},
the paper highlights the advancements made by traditional texture feature analysis and deep learning's capacity to discern complex patterns from large datasets.
It acknowledges the enduring challenges of data annotation and the heterogeneity of data sources, advocating for innovative solutions to enhance data quality and model performance.
\textit{For the needle detection},
traditional geometrical detection methods and DL's automatic pattern recognition have improved accuracy and efficiency.
This paper points out the necessity for more robust techniques to address poor imaging quality and reduce annotation errors, suggesting the exploration of weakly-supervised and semi-supervised learning methods as future directions.
In conclusion, this survey provides an exhaustive overview of the current methodologies and challenges in prostate TRUS image processing and suggests potential research directions to advance the field further.

\section*{Conflict of interest}
The authors declare no competing interests.

\section*{Acknowledgments}
This work was supported in part by the National Natural Science Foundation of China under Grants 62471306 and 62071305,
in part by the Guangdong-Hong Kong Joint Funding for Technology and Innovation (2023A0505010021),
and in part by the Guangdong Basic and Applied Basic Research Foundation under Grant 2022A1515011241.

\bibliographystyle{apalike}

\bibliography{cas-refs}


\end{document}